\documentclass[fleqn,usenatbib]{mnras}

\usepackage{newtxtext,newtxmath}
\usepackage{threeparttable}
\usepackage{lipsum}
\usepackage{xcolor}

\usepackage[T1]{fontenc}

\DeclareRobustCommand{\VAN}[3]{#2}
\let\VANthebibliography\thebibliography
\def\thebibliography{\DeclareRobustCommand{\VAN}[3]{##3}\VANthebibliography}

\usepackage{graphicx}	
\usepackage{amsmath}	
\usepackage{tabularx}

\usepackage{booktabs}
\usepackage{float}

\newcommand{\dechms}[4]{$#1^{\rm h}#2^{\rm m}#3\mbox{$^{\rm s}\mskip-7.6mu.\,$}#4$}

\newcommand{\decdms}[4]{$#1^{\circ}#2'#3\mbox{$''\mskip-7.6mu.\,$}#4$}
\newcommand{\msec}[2]{$#1\mbox{$''\mskip-7.6mu.\,$}#2$}
\newcommand{\Lynds}{L\,1551~IRS\,5 }
\newcommand{\LyndsP}{L\,1551~IRS\,5}
\newcommand{\COgas}{$\textup{C}^{18}\textup{O} \: (2-1)$ }
\newcommand{\COgasP}{$\textup{C}^{18}\textup{O} \: (2-1)$}

\newcommand{\Msun}{M$_\odot$}

\title[\Lynds structure]{FAUST XXVII: The circumbinary disk and the outflow of the \Lynds binary system}

\author[A. Durán et al.]{
Aurora Durán,$^{1}$\thanks{E-mail: a.duran@irya.unam.mx}
Laurent Loinard,$^{1,2,3}$
Pedro R. Rivera-Ortiz,$^{1}$
Geovanni Cortés-Rangel,$^{1}$
Eleonora Bianchi,$^{5}$\newauthor
Paola Caselli, $^{7}$
Cecilia Ceccarelli,$^{4}$
Claire J. Chandler,$^{16}$
Claudio Codella,$^{5}$
Nicolás Cuello,$^{4}$\newauthor
Marta De Simone,$^{6,5}$ 
Tomoyuki Hanawa,$^{13}$
Doug Johnstone,$^{14,15}$
François Menard,$^{4}$
Maria José Maureira,$^{7}$\newauthor
Anna Miotello,$^{6}$
Linda Podio,$^{5}$
Takeshi Sakai,$^{8}$
Giovanni Sabatini,$^{5}$
Leonardo Testi,$^{9,5}$
Charlotte Vastel,$^{17}$\newauthor
Ziwei Zhang,$^{10}$
Nami Sakai,$^{5}$
Satoshi Yamamoto,$^{11,12}$
\\
$^{1}$Instituto de Radioastronomía y Astrofísica, Universidad Nacional Autónoma de México Apartado Postal 3-72, 58090 Morelia, Michoacán, México\\
$^{2}$Black Hole Initiative at Harvard University, 20 Garden Street, Cambridge, MA 02138, USA\\
$^{3}$David Rockefeller Center for Latin American Studies, Harvard University, 1730 Cambridge Street, Cambridge, MA 02138, USA\\
$^{4}$Univ. Grenoble Alpes, CNRS, IPAG, 38000 Grenoble, France\\
$^{5}$INAF, Osservatorio Astrofisico di Arcetri, Largo E. Fermi 5, I-50125, Firenze, Italy\\
$^{6}$ESO, Karl Schwarzchild Srt. 2, 85478 Garching bei Munchen, Germany\\
$^{7}$Center for Astrochemical Studies, Max-Planck-Institut für extraterrestrische Physik (MPE), Gießenbachstr. 1, D-85741 Garching, Germany\\
$^{8}$Graduate School of Informatics and Engineering, The University of Electro-Communications, Chofu, Tokyo 182-8585, Japan\\
$^{9}$Dipartimento di Fisica e Astronomia ``Augusto Righi'' Viale Berti Pichat 6/2, Bologna, Italy\\
$^{10}$RIKEN Cluster for Pioneering Research, 2-1, Hirosawa, Wako-shi, Saitama 351-0198, Japan\\
$^{11}$SOKENDAI (The Graduate University for Advanced Studies), Hayama-cho, Miura-gun, Kanagawa 240-0193, Japan\\
$^{12}$Research Center for the Early Universe, The University of Tokyo, 7-3-1, Hongo, Bunkyo-ku, Tokyo 113-0033, Japan\\
$^{13}$Center for Frontier Science, Chiba University, 1-33 Yayoi-cho, Inage-ku, Chiba 263-8522, Japan\\
$^{14}$NRC Herzberg Astronomy and Astrophysics, 5071 West Saanich Road, Victoria, BC, V9E 2E7, Canada\\
$^{15}$Department of Physics and Astronomy, University of Victoria, Victoria, BC, V8P 5C2, Canada\\
$^{16}$National Radio Astronomy Observatory, PO Box O, Socorro, NM 87801, USA\\
$^{17}$IRAP, Université de Toulouse, CNRS, CNES, UPS, Toulouse, France
}

\date{Accepted XXX. Received YYY; in original form ZZZ}

\pubyear{2024}

\begin{document}
\label{firstpage}
\pagerange{\pageref{firstpage}--\pageref{lastpage}}
\maketitle

\begin{abstract}

Using continuum and $\textup{C}^{18}\textup{O}\:(2-1)$ line data obtained from the large ALMA program FAUST, we studied the structure of the protostellar binary system \Lynds at scales between 30 and 3,000 au to constrain its properties, from the circumstellar and circumbinary disks up to the envelope and outflow scales, which exhibits complex and entangled structures at the scales of its inner and outer envelopes, presumably caused by the influence of the central binary. Assuming a dust-to-gas
ratio of 100,  we calculated the  dust+gas mass for the circumbinary disk and each circumstellar disk of the binary,  obtaining 0.018 \Msun\, for the circumbinary disk, 0.004 \Msun\, and 0.002 \Msun\, for the northern and southern circumstellar disk respectively.
From the line emission, we retrieved the gas masses for each structure component. 
With the $\textup{C}^{18}\textup{O}\:(2-1)$ PV diagram along the circumbinary disk, we were able to constrain the centrifugal barrier, ${r_{CB}=55}$ au, update the specific angular momentum, ${j\sim270}$~au~km~s$^{-1}$. We built an analytical model that can be used to predict the influence of the morphology of the outflow and a few dynamic features that can reproduce the system emission, allowing us to explain and discern the outflow contribution from the complex emission due to the binary. Additionally, we inferred the density power law index, ${\alpha=1.7}$, and the envelope rotation velocity, $\varv_{c}=2$~km~s$^{-1}$. 
Finally, the observations gave us the physical constraints to obtain a coherent outflow model for \LyndsP.

\end{abstract}




\section{Introduction} \label{sec:intro}



Star formation has been studied for the last half century and there is now an accepted paradigm for the formation process of a low mass single star \citep[e.g.,][]{Shu1987, McKee2007, Frank2014}.
In this scenario, the early formation
 \citep[Class~0 and Class~I, age $\leq 10^{5}$ yr,][]{Andre1999, Andre2000}  stages of a Solar-type star are characterized by a disk/envelope system that can be divided into three zones: envelope, barrier, and disk \citep[e.g.][]{Oya2016},
at scales less than 1000 au. At larger scales, these Class~0 and Class~I objects also show jet-driven outflows \citep{Frank2014}.
 However, it is well known that a large fraction of stars belong to binary or multiple systems and that high levels of multiplicity are already established during the star-formation process \citep[e.g.][]{Duchene2007, Reipurth2014}. The two main theories to explain the formation of multiple systems are the fragmentation of a gravitationally unstable circumstellar disk \citep[e.g.][]{Adams1989} and the turbulent fragmentation of the molecular cloud \citep[e.g.][]{Padoan2007, Goodwin2004}. The first scenario is expected to form binaries with a separation of less than 100 astronomical units while the latter will create more widely separated multiple systems \citep{Tobin2016}. 
To gain better insights into the formation of multiple and binary stars, it is essential to characterize the architectures of these systems in their early stages.

\Lynds is often considered a prototype of a binary system formed by the fragmentation of a gravitationally unstable circumstellar disk \citep{Lim2016}. It has been amply studied during the last decades, ever since its first detections at radio wavelengths \citep[e.g.][]{Snell1980,Rodriguez1986}. These studies have revealed various types of structures, characterized by specific kinematic properties at different scales 
\citep[e.g.][]{Momose1998, Chou2014, Lim2016, Feeney-Johansson2023}.
\Lynds was shown to be a binary system by \citet{Looney1997}; it is located in the Taurus Molecular Cloud at a distance of 146.4 $\pm$ 0.5 pc \citep{Galli2019}, and it has a bolometric luminosity 
$L_{bol}= 30 - 40$ L$_{\sun}$ \citep{Liseau2005}.
Due to its optical spectrum seen via a reflection nebula, it is classified as a FUor-like object \citep{Mundt1985,Connelley2018},
and it is considered a prototypical Class I source \citep{Adams1987}. 
This system is composed of a northern (N) and a southern (S) protostar,
of $\sim$ 0.8 M$_{\sun}$ and $\sim$ 0.3 M$_{\sun}$, respectively, according to \citet{Liseau2005}. These components are separated by   \msec{0}{36} that, given the distance to \LyndsP, corresponds to $\sim 50$ au \citep{Lim2016}. The orbital period of the system has been previously estimated to be $ \simeq $ 260 yr \citep{Rodriguez2003a}. Recently,  \citet{HernandezGarnica2024} combined all existing ALMA and VLA observations of \Lynds\ to refine the orbital parameters. They find an orbit consistent with coplanarity with the circumbinary disk, a total mass (N+S) of $0.96\pm0.17$ \Msun, and an orbital period of 300 years.

Additionally, \Lynds is associated with two powerful nearly parallel radio jets \citep{Rodriguez2003b}. Recently \citet{Feeney-Johansson2023} showed that there is significant temporal  variation in the ionized mass-loss rate of the jets, particularly for the N component. This implies a variation in the radio flux density by a factor of $\sim$ 5 from epoch to epoch, although \citet{Park2021} did not detect any strong indication of mid-IR variability from this source.
Each individual protostar has a circumstellar disk, and the system is surrounded by a circumbinary disk with a mass of 0.02 - 0.03 M$_{\sun}$ \citep{Lim2016}.
\citet{Bianchi2020} provided evidence of a hot corino in the north source and they showed the chemical richness of Class I protostars on a Solar system scale. So far, the components such as disks, envelope, outflow, and jet have been studied individually. 
Given the outflow size and the separation of the individual components we expect that there must be an interaction between these structures and the overall architecture of the system. Thus, in this paper, we look for evidence of the interaction in the inner part of the envelope and the dynamical outflow. 
To constrain the properties of the system, from the circumbinary disk to the scale of the outflow, we use data obtained from the ALMA (Atacama Large Millimeter/submillimeter Array) Large Program FAUST (Fifty AU STudy of the chemistry in the disk/envelope system of solar-like protostars 
2018.1.01205.L, PI: S. Yamamoto; \citet{Codella2021}; 
\url{http://faust-alma.riken.jp}).



In this paper, we first present our Band 6 observations of \Lynds obtained using several ALMA configurations (section \ref{sec: Observations}). Then, in section \ref{sec: Cont_Emission}, we present and discuss the continuum observations and obtain the physical parameters of the circumbinary disk, while in section \ref{sec: Maps} we describe the morphology of the \COgas at envelope scales. The kinematics are discussed in section \ref{sec: Kynematics}. Finally, we summarize and conclude in section \ref{sec: Conclusions}.

\section{Observations and Data Reduction} 
\label{sec: Observations}

The observations were obtained with ALMA through the FAUST Large Program in Band 6.
We used three different configurations: the 7~m array of the Atacama Compact Array (ACA/Morita Array), and the 12~m array in C43-5 for an extended configuration, and C43-2 for a compact configuration. The observation parameters are listed  in Table \ref{table:parameters}. The observations were centered at $\alpha_{J2000} = $ \dechms{04}{31}{34}{14},
$\delta_{J2000} = $ +\decdms{18}{08}{05}{10}.

\begin{table*}
\caption{Observation Parameters}
\centering
 \begin{threeparttable}[b]
\begin{tabular}{@{}llll@{}}
\toprule \toprule
        Parameter                         & ACA              & C43-2       & C43-5       \\ 
\midrule
        Observation Date                  & 2018 Oct 22, 23  & 2019 Jan 06   & 2018 Oct 25 \\
        Time on source (min)              & 42.9             & 12.63         & 51.25       \\
        Antennas                          & 12, 11 \tnote{a} & 51            & 47          \\
        Primary beamwidth (arcsec)        & 45.8             & 26.7          & 26.7        \\
        Total Bandwidth (GHz)             & 0.062            & 0.059         & 0.059          \\
        Continuum Bandwidth (GHz)         & 2.000            & 1.875         & 1.875          \\
        Proj. baseline range (m)          & 7.2 - 39.4       & 12.0 - 437.9  & 12.0 - 1169.45 \\
        Bandpass calibrator               & J0423-0120       & J0440+1437    & J0423-0120  \\
         Flux calibratior                  & J0423-0120       & J0440+1437    & J0423-0120  \\
        Gain calibrator                    & J0510+1800       & J0510+1800    & J0510+1800  \\
 Resolution (arcsec) (P. A. (deg))\tnote{b} & 6.3 $\times$ 4.5 (-79.2) & 0.67 $\times$ 0.63 (21.9) & 0.27 $\times$ 0.25 (65.9) \\
 rms (mJy beam$^{-1}$ channel$^{-1}$)\tnote{b} & 4.3        & 0.8             & 0.19   \\
 Resolution (arcsec) (P. A. (deg))\tnote{c}    & 7.4 $\times$ 5.7 (-82.3) & 1.07 $\times$ 0.99 (50.1) & 0.49 $\times$ 0.39 (29.5) \\
rms (mJy beam$^{-1}$ channel$^{-1}$)\tnote{c}  & 5.1        & 0.56            & 0.61  \\
 Resolution (arcsec) (P. A. (deg))\tnote{d}    & 6.6 $\times$ 5.1 (-81.7) & 0.39 $\times$ 0.36 (41.0) & 0.36 $\times$ 0.31 (35.3) \\
 rms (mJy beam$^{-1}$ channel$^{-1}$)\tnote{d} & 3.8        & 0.23            & 0.19 \\
\bottomrule 
\end{tabular}
\begin{tablenotes}
       \item [a] 12 antennas for Oct 22, and 11 antennas for Oct 23.
       \item [b] Uniform weighting (robust, r = --2).
       \item [c] Natural weighting (r = 2).
       \item [d] Intermediate weighting (r = 0.5).
     \end{tablenotes}
\end{threeparttable}     
\label{table:parameters}
\end{table*}



 The data were reduced with CASA, the Common Astronomy Software Applications for Radio Astronomy \citep{The_CASA_Team_2022} using a modified version of the ALMA calibration pipeline version 5.6.1-8.el7 and an additional routine to correct for the Tsys and spectral line data normalization\footnote{https://help.almascience.org/kb/articles/what-errors-could-originate-from-the-correlator-spectral-normalization-and-tsys-calibration}.
 A self-calibration procedure using continuum models obtained from line-free channels was employed to align the data from the multiple execution blocks in both position and amplitude, taking care to ensure the models were as complete as possible to avoid impacting the overall flux density scale. The resulting gain tables were then applied back to all the channels, and the final continuum models were the subtracted from the visibility data to form continuum-subtracted line data. 
 All the images were made with the command ``tclean'' in CASA. For the continuum, we produced three images that only use the 12-m antenna array in the  C43-5 configuration. The first one uses a natural weighting scheme of the visibilities to optimize the noise level (Figure \ref{fig:continuum}.a). The second uses uniform weighting to produce a higher resolution version (Figure \ref{fig:continuum}.b), while the last one uses super-uniform weighting. In this last case, we followed a somewhat non-standard cleaning procedure. The cleaning process considered only components associated with the individual protostars \Lynds N and S, and was stopped after they no
 longer appeared in the residual map. The image shown in Figure \ref{fig:continuum}.c only contains the residuals after this interrupted CLEANing process,
 and provides a clear view of the circumbinary disk alone. For the \COgas line, we combined all observations (ACA, C43-2, and C43-5) and used Briggs weighting with a robust parameter set to 0.5 
 {to have a balance between noise and resolution, since we were looking for the extended structure}.
 Each velocity channel was cleaned separately using the ``automultithresh'' option of tclean.

\section{Continuum Emission}
\label{sec: Cont_Emission}

In this section, we present and discuss the maps of the continuum emission of \Lynds in the frequency range from 216.112 GHz to 233.795 GHz, using the data from the C43-5 array. With these observations and using different imaging techniques, we can retrieve structures at multiple scales, from a few tens of au up to a few thousand au. We observe an extended emission believed to be part of the envelope of the source; additionally, we see the individual circumstellar disks (northern and southern) surrounded by the circumbinary disk (CBD) which we could isolate using the non-standard interrupted CLEAN strategy described above.

\begin{figure}
\centering
   \includegraphics[width=0.8\columnwidth]{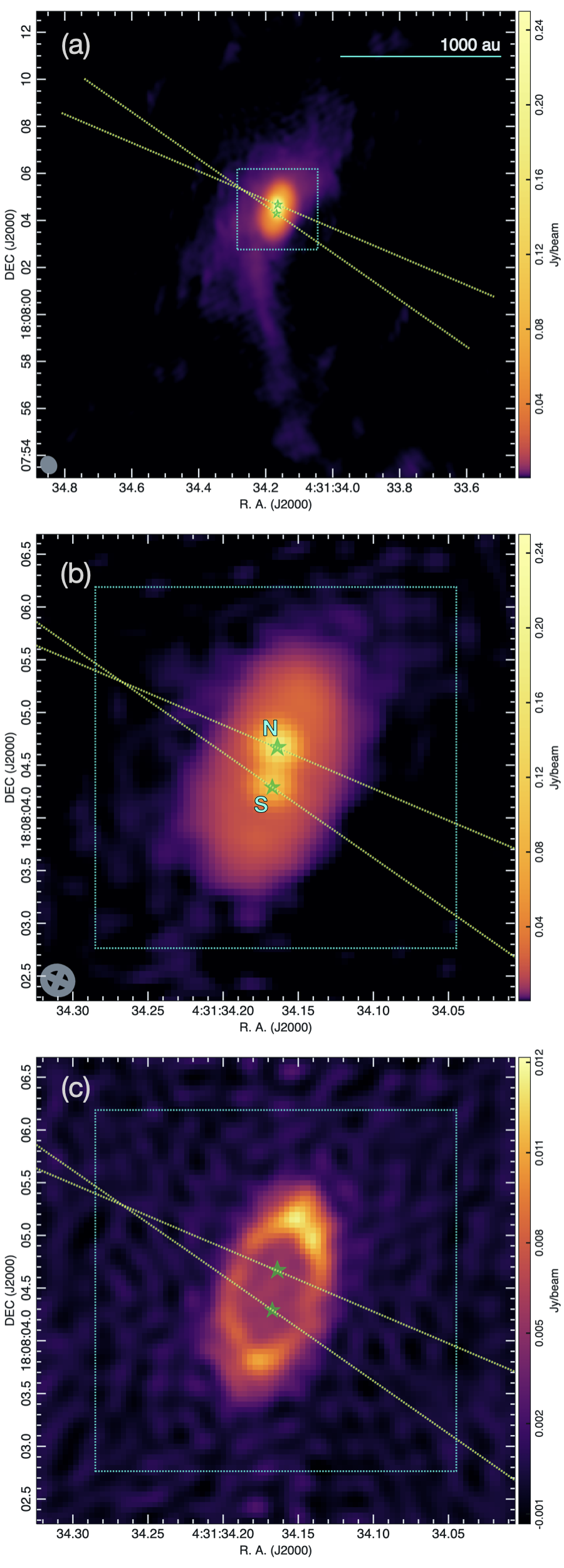}
   \caption{Panel (a) illustrates \Lynds across a broad field of view. In addition to the prominent central source, a faint and elongated structure can be discerned towards the southern region of the source. Panel (b) presents a close-up view of the source \Lynds revealing its internal components, such as the circumbinary disk, and the circumstellar disks of the northern (N) and southern (S) sources. The dashed square is the same size in all figures, 500au$\times$500au, and it is used to represent the same scale. Finally, Panel (c) shows the CBD removing CN and CS revealing a notable asymmetry in the dust emission of the CBD; its northern portion appears brighter than its southern portion. The dashed square has the 
   same size in all panels.
   The dashed yellow lines represent an extension of the jets for each source (N and S) reported by \citet{Rodriguez2003b}.
   }
   \label{fig:continuum}
\end{figure}

To study the extended emission we use the image obtained with natural weighting. Figure \ref{fig:continuum}.a shows the full field of view of that image where emission is detected. Most of the emission is contained inside the dashed square at the center of the panel (at scales of a few hundred au) but we can also distinguish a broad band of diffuse emission at a scale of about 1000 au oriented at a position angle of about $-30^\circ$. This is likely associated with the flattened inner envelope of \LyndsP. Connecting to and extending toward the south-southwest of this structure is a filament of emission with an approximate length of 1000 au. This structure is most likely associated with the outflow cavity edge formed by the powerful radio jets reported by \citet{Rodriguez2003b}. 

We used the uniform weighting image (Figure \ref{fig:continuum}.b) to zoom in on the central compact component. At the corresponding scales of a few hundred au, the emission is dominated by the circumstellar and circumbinary disks. The CBD in \Lynds\ has a size similar to that of comparable structures in other sources, around 300 au,
\citep[e.g.,][]{Maureira2020,Vastel2022,Codella2024}. The circumstellar disks of the northern and the southern protostellar sources (CN and CS respectively) remain unresolved at our angular resolution. Figure \ref{fig:continuum}.c zooms in further to show the structure of the CBD (note that the dashed square has the same size in all the panels of Figure \ref{fig:continuum}). As mentioned earlier, the individual sources (CN and CS) have been removed from this image. 
Though this image is partially artificial due to the removal of the central emission during the imaging process, it gives us a clearer view of the CBD alone. A clear brightness asymmetry is visible in the dust emission of the CBD of \Lynds, with the northern side significantly brighter than the southern side. This asymmetry can be interpreted as a result of tidal forces created by the central binary system (Cuello et al., in prep) and is out of the scope of this work.


From the continuum emission, it is possible to obtain some physical parameters of the CBD. In particular, we measure its size and orientation to be:
\begin{itemize}
    \item $\theta_{max}$ = 1.8245 $\pm$ 0.0075 arcsec $\equiv$ 267.2 $\pm$ 2.2 au,
    \item $\theta_{min}$ = 0.8921 $\pm$ 0.0037 arcsec $\equiv$ 130.6 $\pm$ 1.2 au,
    \item P.A. = 160.9$^\circ$ $\pm$ 0.2$^\circ$,
\end{itemize}
where $\theta_{max}$ and $\theta_{min}$ are the major and minor {semi-axes} of the CBD, respectively, {considering a half power emission.}
Assuming that the CBD is intrinsically circular, $\theta_{min}$ must be equal to  
${ \theta_{max}\cos i}$ where $i$ is the inclination of the CBD with respect to the plane of the sky. We obtain $i$ = 60.73$^\circ$ $\pm$ 0.14$^\circ$.

The dust masses of the circumbinary disk and the circumstellar disks can be calculated using the method described by \citep{Bergin2017}:
\begin{equation}
    M_{\text{disk}} = \frac{F_{\nu} d^{2}}{\kappa_{\nu} B_{\nu}(T_{\text{dust}})},
\end{equation}
where $F_{\nu}$ is the total continuum density flux, $d$ is the distance to the source, $\kappa_{\nu} = 2.25$ cm$^{2}$ g$^{-1}$   is the mass absorption coefficient produced by dust grains, and $B_{\nu}(T_{\text{dust}})$ is the Planck function at a dust temperature $T_{\text{dust}}$.  \citet{Bergin2017} assumed a minimum temperature for the dust of 100 K.

We use values for the $F_{\nu}$ of 0.801 Jy for CBD, 0.1416 Jy for CN, and 0.0756 Jy for CS.
Then, assuming a dust-to-gas ratio of 100 \citep{Bohlin1978}, the total disk dust+gas disk masses are:

\begin{itemize}
    \item $M_{CBD} = 0.018 \pm 0.008 $ M$_{\odot}$,
    \item $M_{CN}  = 0.004 \pm 0.001 $ M$_{\odot}$,
    \item $M_{CS}  = 0.002 \pm 0.001 $ M$_{\odot}$,
\end{itemize}
for the circumbinary, northern circumstellar, and southern circumstellar disks, respectively. 
In accordance with the mass lower limits 
reported by \citet{Cruz2019}.
We note that the masses of the circumstellar disks are of the order of $\sim$1\% of the masses of their associated stars (0.8 and 0.3 M$_\odot$, respectively, according to \citealt{Liseau2005}). This remains true for the $\sim$ 20\% lower mass of the system recently determined by \citet{HernandezGarnica2024}.
Though we use a value of 100 for the dust-to-gas ratio, we are aware that the value might impact the calculated mass \citep[e.g.,][]{Ansdell2017,Okuzumi2025}.

{For circumstellar compact disks, 100~K is a reasonable assumption (Maureira et al., in prep.). However, this assumption will need to be revised in detail in the future, once resolved dust temperature measurements for the CBD in \LyndsP. A significantly lower temperature would result in a commensurably higher dust mass. For instance, for $T_{\text{dust}}$ = 35~K, we obtain a mass of $M_{CBD} = 0.064  $ M$_{\odot}$.}


\section{$\textup{C}^{18}\textup{O} \: (2-1)$ line analysis}
\label{sec: Maps}
\subsection{Channel Maps Qualitative Description}
\begin{figure*}
    \centering
    \includegraphics[width=\linewidth]{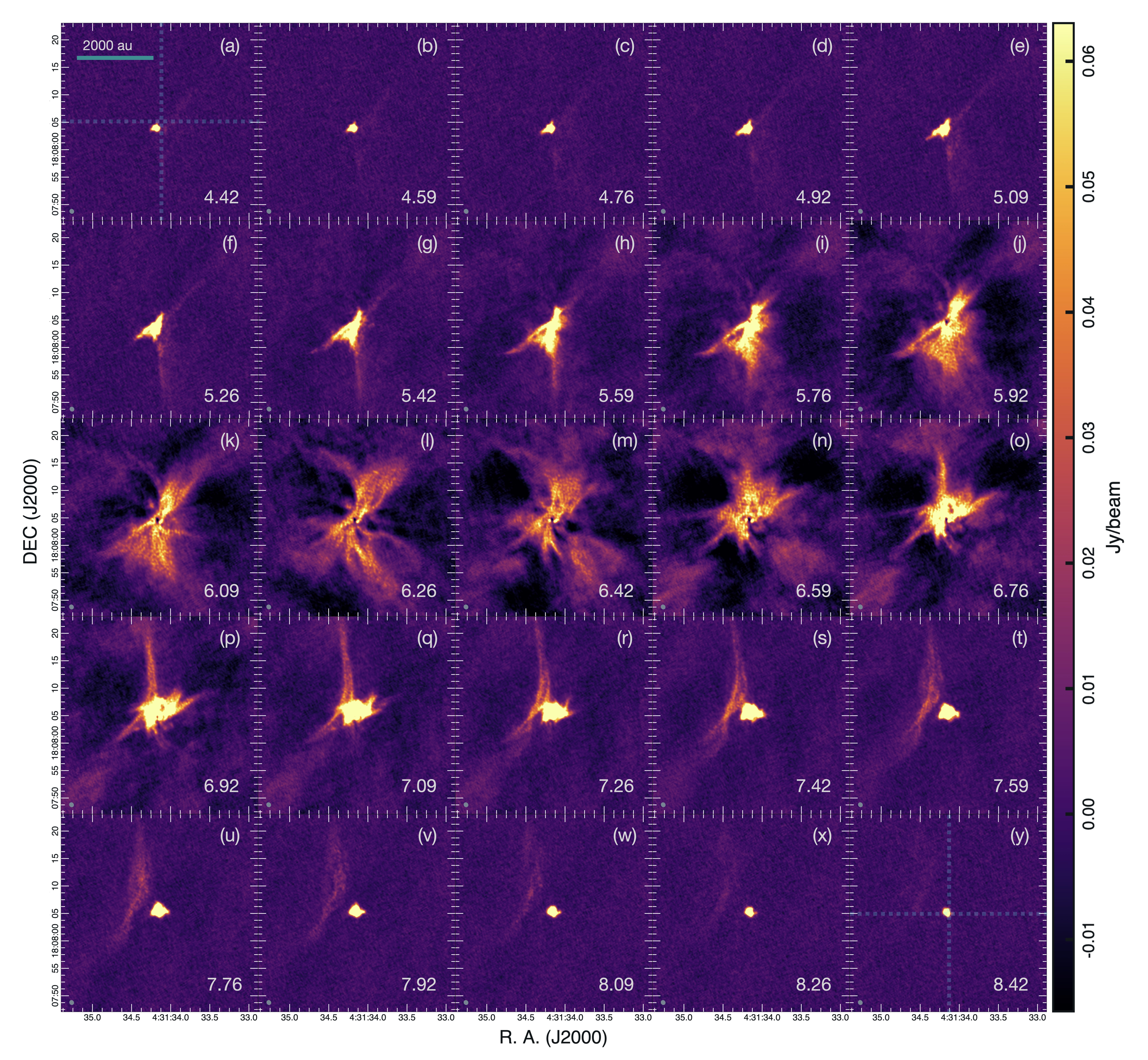}
    \caption{Velocity channel maps of the \COgas line observed in \Lynds. 
    The velocity [km s$^{-1}$] for each channel is shown in the lower right corner. The synthesized beam is shown at the bottom left of each panel. The intensity scale is indicated by the color bar to the right of the figure. The systemic velocity is $\varv_{lsr}=6.45$~km~s$^{-1}$.}
    \label{fig:C18O_channels}
\end{figure*}

As mentioned above, the data from three different array configurations 
 (ACA, C43-2, and C43-5) were combined to produce the \COgas line data cubes. This provides access to 
optimized sensibility at all accessible scales. Given the \COgas line frequency, 219.560 GHz, the median area beam is \msec{0}{42} $\times$ \msec{0}{33}.
In what follows, we will describe the channel maps assuming a systemic velocity of $\varv_{lsr}=6.45$~km~s$^{-1}$, which was obtained from the moment one map that will be presented later in the text. 
We explore the velocity range between $-$9.85 and $+$10.14 ~km~s$^{-1}$, divided in 121 velocity channels, with respect to the systemic velocity. 
In Figure \ref{fig:C18O_channels}, we present 25 channel maps which adequately sample the overall emission; panel (m) samples the emission closest to the systemic velocity.

In most channels, the emission is dominated by arc-like features. Many of these structures trace transversal cuts through the outflow cavity walls, similar to the model for L 1527 presented by \citet{Oya2016}. We can see these arcs on the southern side from panels (e) - (h), and their apex gets closer to the source as the channel velocity approaches the systemic velocity, close to panel (m). Similarly, in the northern case, the apex of the arcs moves away from the source, from panels (t) - (v), as the channel velocity recedes from the systemic velocity. In the central panels, (k) to (o), besides the cavities, we observe diffuse and filamentary emission that could be tracing a mixture of infalling and outflowing material.
The possibility that some of these filaments could be accretion streamers {\citep{Pineda2023}} would need to be examined carefully using, in particular, kinematics information.

At the systemic velocity itself, panel (m), we see the most extended structure in \COgas centered on the binary source. This structure traces an X-shaped pattern with a symmetry axis parallel to the emitted jets.


\begin{figure}
    \centering
    \includegraphics[width=1\linewidth]{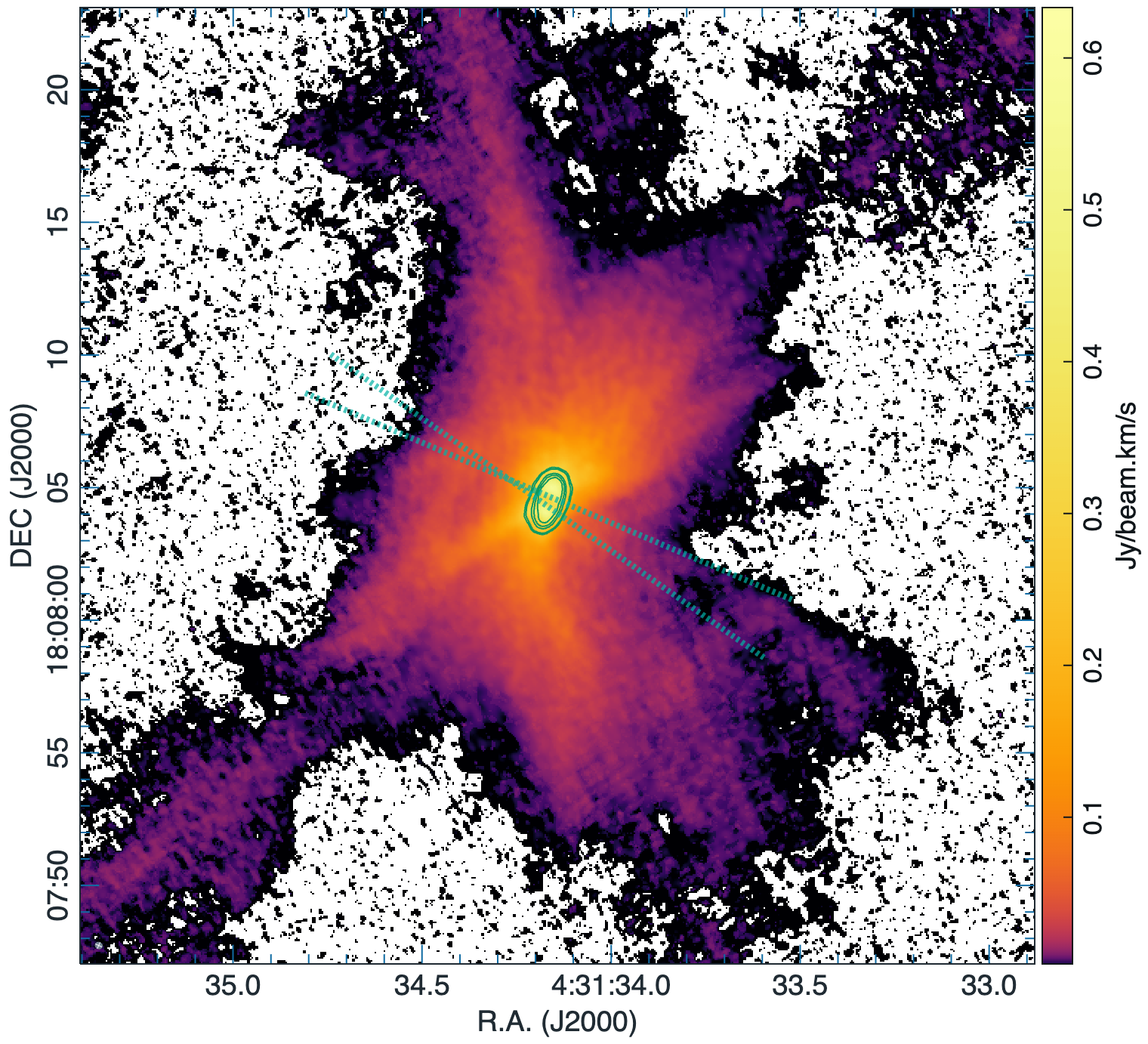}
    \caption{\COgas moment zero map towards \Lynds. The circumbinary disk is shown in contours at the image center. The contour levels are 15, 20, 50, 70, 100 $\times \sigma_{rms}$, where $\sigma_{rms}$= 0.16 mJy beam $^{-1}$. The dashed green lines represent the extension of the jets reported by \citet{Rodriguez2003b}. 
    }
    \label{fig:MomZero}
\end{figure}

Figure \ref{fig:MomZero} shows the moment zero of the \COgas line emission of \LyndsP, which is the line integrated intensity. The circumbinary disk is shown in contours at the image center. The directions of the two jets reported by \citet{Rodriguez2003b} are shown in green dashed lines. The jets have position angles of 67$^{\circ}$ $\pm$ 3$^{\circ}$ and 55$^{\circ}$ $\pm$ 1$^{\circ}$, for the north and south sources, respectively, and are approximately perpendicular to the CBD main axis. In addition, the moment zero map shows thin strands, which we had already noticed in the channel maps of figure \ref{fig:C18O_channels}, that when integrated together make a thick wall. In particular, we can now clearly see a large outflow opening angle. We note, however, that we do not detect \COgas line along the jets themselves. 

\begin{figure}
    \centering
    \includegraphics[width=1\linewidth]{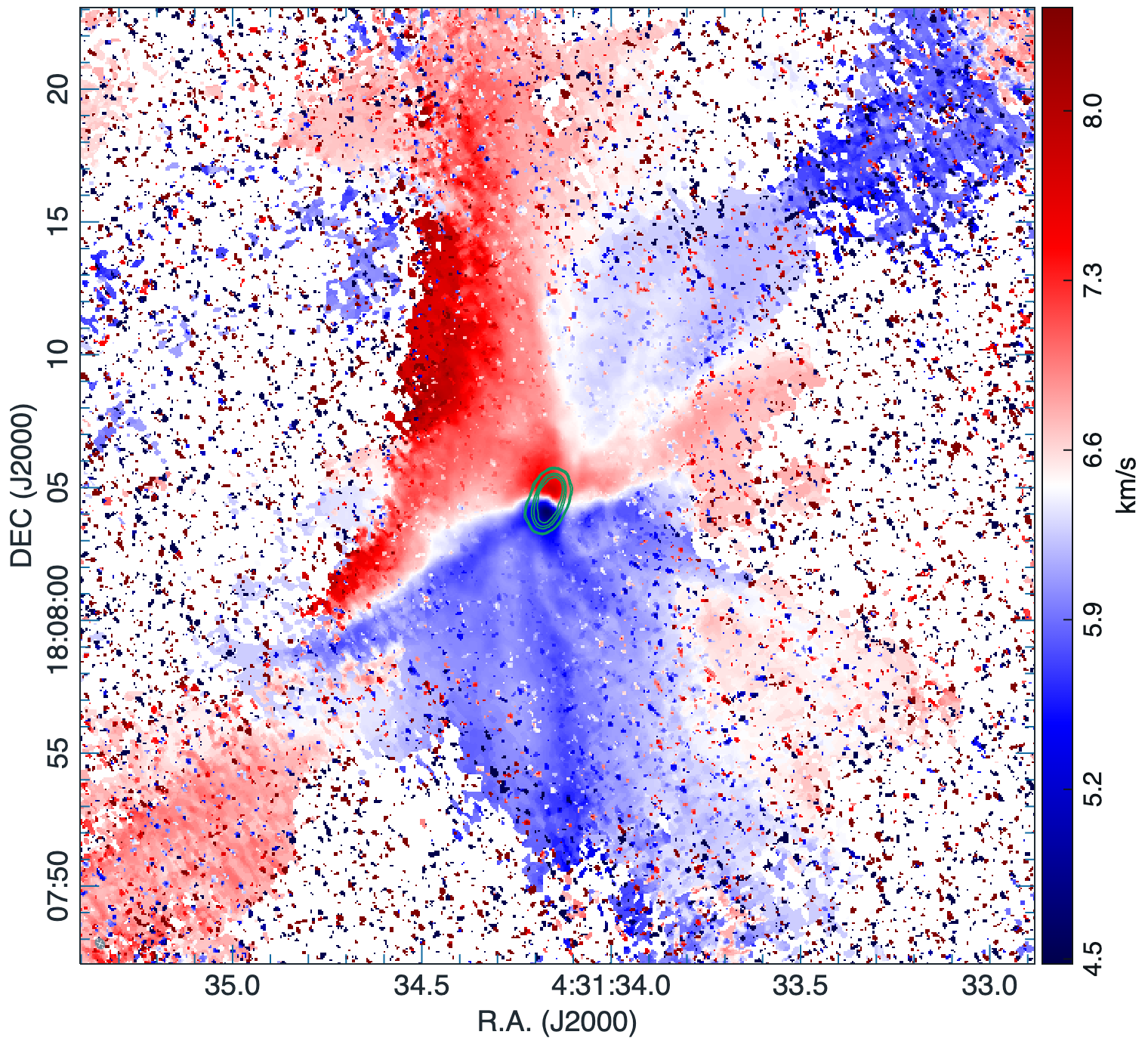}
    \caption{\COgas moment one map towards \Lynds. The circumbinary disk is shown in contours at the image center. The contours levels are the same as in figure \ref{fig:MomZero}.
    Most of the redshifted material corresponds to the northern cavity while the blueshifted material is found in the southern cavity. 
     There is a extended emission between the cavities in the direction of the disk main axis.}
    \label{fig:MomOne}
\end{figure}

In figure \ref{fig:MomOne} we present the \COgas moment one map tracing the velocity field. Gas to the southeast of the protostars appears to be blueshifted while gas to the northwest is redshifted with respect to the systemic velocity. This confirms, for the first time in \LyndsP,  a co-rotation of the envelope and the CBD. We can also see that inside the CBD the velocities are larger than in the envelopes, as expected from angular momentum conservation. In this figure, we also note the X-shape traced at envelope scales. It matches spatially with the X-shape also shown in figure \ref{fig:MomZero}. 
Overall, the structure of the moment one map reflects two processes: on the one hand, the rotation of the disk/inner envelope system and, on the other, the expulsion of material along the jet axes.

\subsection{Gas masses obtained from the \COgas line}
We obtained the observational parameters from the \COgas images using a Gaussian fit (in CASA) to the emission structures detected in the data cube (see Table \ref{pbp}). We used these parameters to calculate gas mass from the line emission of \COgas following the methodology described by \citet{Cortes-Rangel2020}. We calculated the mass associated with each source structure (see Table \ref{table:Mass}). 
We use an excitation temperature T$_{ex}=20$ K, typical for this kind of regions, and X$_r \sim 5 \times$ 10$^{6}$ for the ratio of H$_2$ abundance compare to C$^{18}$O  \citep{AntoineRoueff2021} 
and a distance to the source of 146 pc. We estimated the amount $\int I_{\nu} d\varv$ from the fit to the spectrum of the detected emission; here $I_{\nu}$ is the average intensity and d$\varv$ is the channel width in the velocity range emission, described in table \ref{table:Mass}.
We note the gas masses are about an order of magnitude lower than the mass estimates given in section \ref{sec: Cont_Emission}. 
This discrepancy could be explained by a value of dust-to-gas ratio different than 100, since there is a large uncertainty on this value. According to the observations in \citet{Ansdell2017}, this ratio depends on the age of the disk, and can be as low as 10. This same conclusion has been reached on theoretical grounds by \citet{Okuzumi2025}. In the case of \LyndsP, M.J.\ Maureira et al, (2025, in prep.), shows that the dust emission at 1.3 mm is optically thick in the inner regions. This implies that the radiation coming out at the frequency of the \COgas\ will be optically thick as well, as the opacity there will be the sum of the dust and line opacities and will necessarily be larger than the dust opacity by itself. Thus, masses derived from the \COgas\ line will be underestimated even if the gas-to-dust ratio had a standard value of 100.

\begin{table}
    \centering
        \caption{Physical observational parameters. The rms was 2 mJy beam$^{-1}$.}
    \begin{tabular}{cccc}
    \hline
\hline
         Structure&  Integer Flux&  Peak Flux&\\
         &  [mJy]&  [mJy beam$^{-1}$]& \\
         \hline
\hline
         CN&  209.2 $\pm$ 1.8&  132.27 $\pm$ 0.74& \\
 CS& 164.5$\pm$ 1.2 & 108.65 $\pm$ 0.53&\\
         CBD (Northern)&25.8 $\pm$ 4&19.2 $\pm$ 1.9& \\
         CBD (Southern)&30.7 $\pm$ 2.8&25.1 $\pm$ 1.4& \\
         Inner envelope (Northern)&3730 $\pm$ 190&205.2  $\pm$ 9.9& \\
         Inner envelope (Southern)&2640 $\pm$ 210&204 $\pm$ 15& \\
         Outer envelope & 34.98 $\pm$ 0.82  & 46.20 $\pm$ 1.10 & \\
         \hline
\hline
    \end{tabular}
    \label{pbp}
\end{table}

\begin{table}
    \centering
        \caption{Gas masses estimated from the line emission of \COgasP. }
    \begin{threeparttable}[b]
    \begin{tabular}{cccl}
    \hline
\hline
         Structure&Range& Size\tnote{a}& Mass \\
         &[km s$^{-1}$]&  [arcsec$^2$]&[$10^{-4}$M$_{\odot}$]\\
         \hline
\hline
 CN & 2.81  \textendash \,
 3.81& 0.20& 2.45\\
 CS& -3.53 \textendash \,
 -2.83& 0.21&1.41 \\
         
         CBD 
(Northern)&2.0   \textendash \,
 3.8&  0.18& 0.49\\
         CBD (Southern)&-4.19  \textendash \,
 -1.99&  1.30&5.14 \\
 Inner envelope (Northern)&0.81  \textendash \,
 1.41& 2.57& 370.95\\

 Inner envelope (Southern)& -1.52  \textendash \,
 -0.86& 1.83&192.06\\
 Outer envelope&-0.36  \textendash \,
 0.34& 41.74& 59.92\\
 \hline
         \hline
    \end{tabular}
\begin{tablenotes}
       \item [a] The size was obtained from gaussian fit to the source/structure emission.

     \end{tablenotes}
     \end{threeparttable} 

    \label{table:Mass}
\end{table}

\subsection{Kinematics}
\label{sec: Kynematics}

\begin{figure}
    \centering
    \includegraphics[width=1\linewidth]{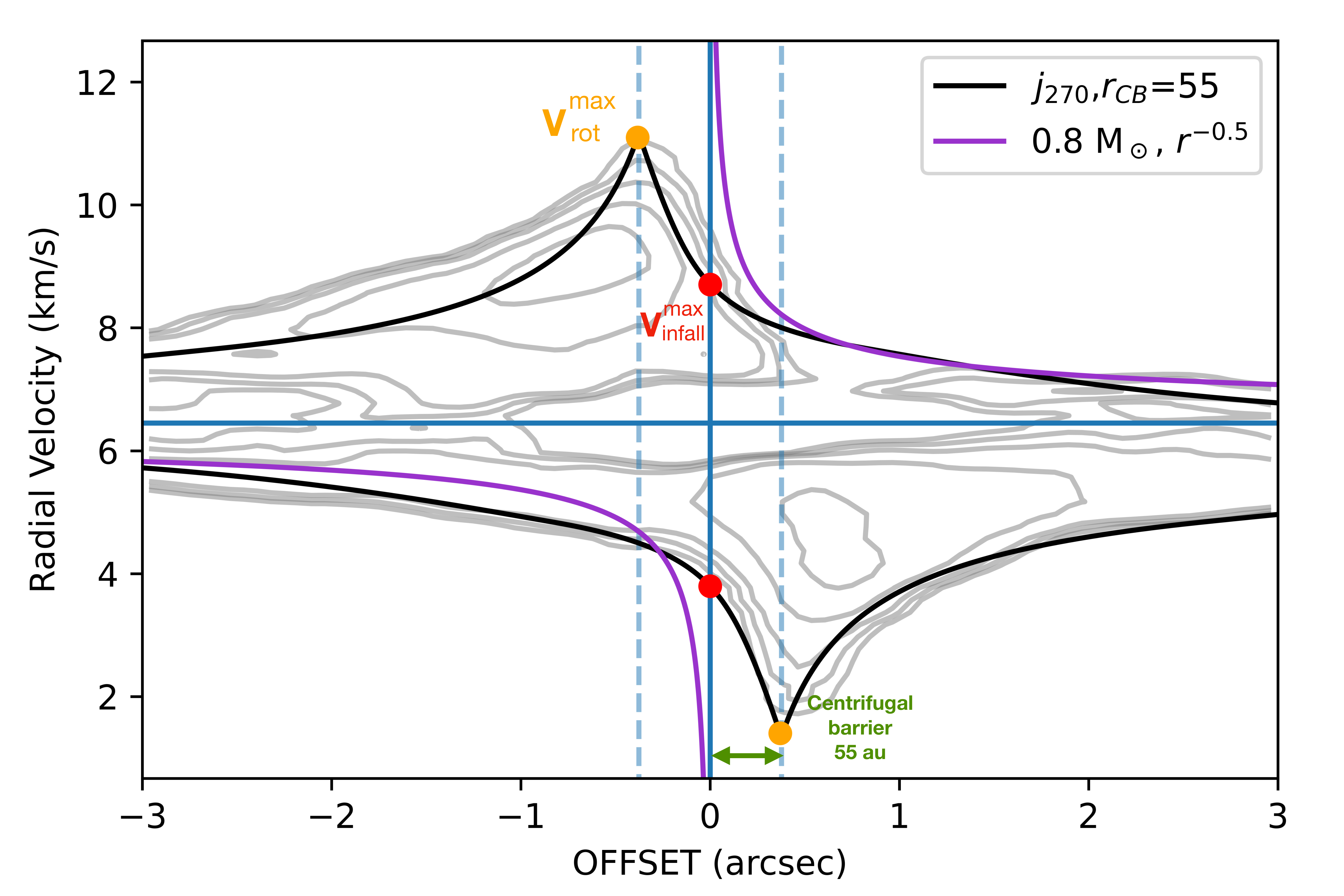}
    \caption{Position-Velocity diagram of the \COgas line emission along the circumbinary disk major axis. 
    Positions are in arcsec relative to the location of the center of the binary component, in the direction of the southern side.
Contours are 3 $\sigma$, 10$\sigma$, 20$\sigma$, 50$\sigma$, 100$\sigma$,  where $\sigma$= 1.73 mJy beam$^{-1}$.
    The systemic velocity at 6.45 km s$^{-1}$ is denoted by a horizontal line. Also, we have superimposed the rotation curves corresponding to a 0.8 system solar mass following a velocity law that decreases inversely with the square root of distance (in purple, keplerian case) and the velocity of a particle along the line of sight (black line) using the kinematic disk model from \citet{Sakai2014,Oya2022}. }
    \label{fig:PVdisk}
\end{figure}

In order to constrain the gas kinematics, we studied the \COgas gas velocity distribution along the disk major axis by means of the Position-Velocity (PV) diagram shown in Figure \ref{fig:PVdisk}. Positions are in arcsec relative to the location of the northern binary component. Gray contours display the observed velocities. The contours are antisymmetric around the source and a velocity of 6.45 km s$^{-1}$, which we have adopted as the systemic velocity (horizontal line in Figure \ref{fig:PVdisk}). 
Also, we have superimposed the rotation curve corresponding to a 0.8 M$_\odot$ system mass, including both protostars, following a velocity law that decreases inversely with the square root of distance (in purple, Keplerian case).


Clearly, the Keplerian line does not match the observed contours at distances close to the center of the source, and it does not explain the observed counter-velocities. Thus, it is necessary to use a disk model that accounts for rotation and infalling matter effect to explain the PV diagram (black line in figure \ref{fig:PVdisk}).
As described by \citet{Sakai2014} and \citet{Oya2022}, the cartesian velocity $\varv$ of a particle along the radius ($r$), measured from the center of the disk, is:
\begin{equation}
    \varv(x,y) = \varv_{\theta}\frac{x}{r} - \varv_{r}\frac{y}{r},
\end{equation}
where $\varv_{\theta}$ is the rotation velocity,
and $\varv_{r}$ the infall velocity. They are given by:
\begin{equation}
    \varv_{\theta} = \frac{j}{r},
    \label{eq:ang_mom_j}
\end{equation}
and
\begin{equation}
    \varv_{r} = \sqrt{\frac{2GM}{r}- \left( \frac{j}{r} \right)^{2}},
    \label{eq:mass_vel}
\end{equation}
where $G$ is the gravitational constant, $M$ is the central mass, and $j$ is the specific angular momentum of the gas.
The radius where the gravitational force is balanced with the centrifugal force is known as $r_{CR}$ and 
the radius of the centrifugal barrier, where kinetic energy is transformed into rotational motion, is defined by
\begin{equation}
    r_{CB}= \frac{j^{2}}{2GM}.
    \label{eq:rcb}
\end{equation}

It follows that $r_{CR}= 2 r_{CB}$ and, 
at $r = r_{CB}$ the rotation velocity is maximum. Thus, we can read-off the value of $r = r_{CB}$ from Figure \ref{fig:PVdisk}, obtaining $r_{CB} = 55$ au, and a corresponding maximum rotation velocity of ${|\varv^{max}_{rot}| = (4.35 \pm 0.2} )  $  km s$^{-1}$.
 Assuming an inclination angle of $i$ = 60.73$^\circ$ the deprojected maximum velocity is $\varv_{max} = (5 \pm 0.1)$ km s$^{-1}$ and using equation  \ref{eq:ang_mom_j}, we derive a specific angular momentum of $j = \varv^{max}_{rot} r_{CB}/ \sin i = 270 \pm 60$ au km s$^{-1}$, that is slightly larger than the value previously reported by \citet{Momose1998}.

At $r_{CB}$, there is no infall, $\varv_r=0$ and, using equations \ref{eq:ang_mom_j} and \ref{eq:mass_vel}, we can determine the value of the central mass which can be expressed by: 
\begin{equation}
    {M= 5.6\times 10^{-4}{\rm{M_{\sun}}}\left[\frac{\varv_{max}}{ \rm{\,km\, s^{-1}} }\right]^{2} \left[ \frac{r_{CB}  }{\rm{au} }\right].}
\end{equation}
From the values above, we obtain a mass for \Lynds of ${M~=~0.8 \pm 0.2}$ \Msun,  consistent with the value obtained with the figure derived from the orbital motion of the binary system by \citet{HernandezGarnica2024}. 
Though this model fits well the PV diagram, we can see a velocity excess in the northern part (negative offset), we attribute it due to the binary interaction of the source creating a tidal effect in the CBD, the same effect we see in the continuum with a excess of dust in the north part of the CBD. 
There are models that account for the time and spatial evolution of the specific angular momentum, and this could affect its value \citep[e.g.,][]{Stahler1994}, nevertheless, the model proposed by
\citet{Sakai2014} and \citet{Oya2022}
 uses parameters obtained directly from the PV diagram, assuming the specific angular momentum is constant over the disk. This seems a reasonable approximation given the superposition of the model in Fig. \ref{fig:PVdisk}.

\subsection{Analytical model} 

\begin{figure}
    \centering
    \includegraphics[width=1.\columnwidth]{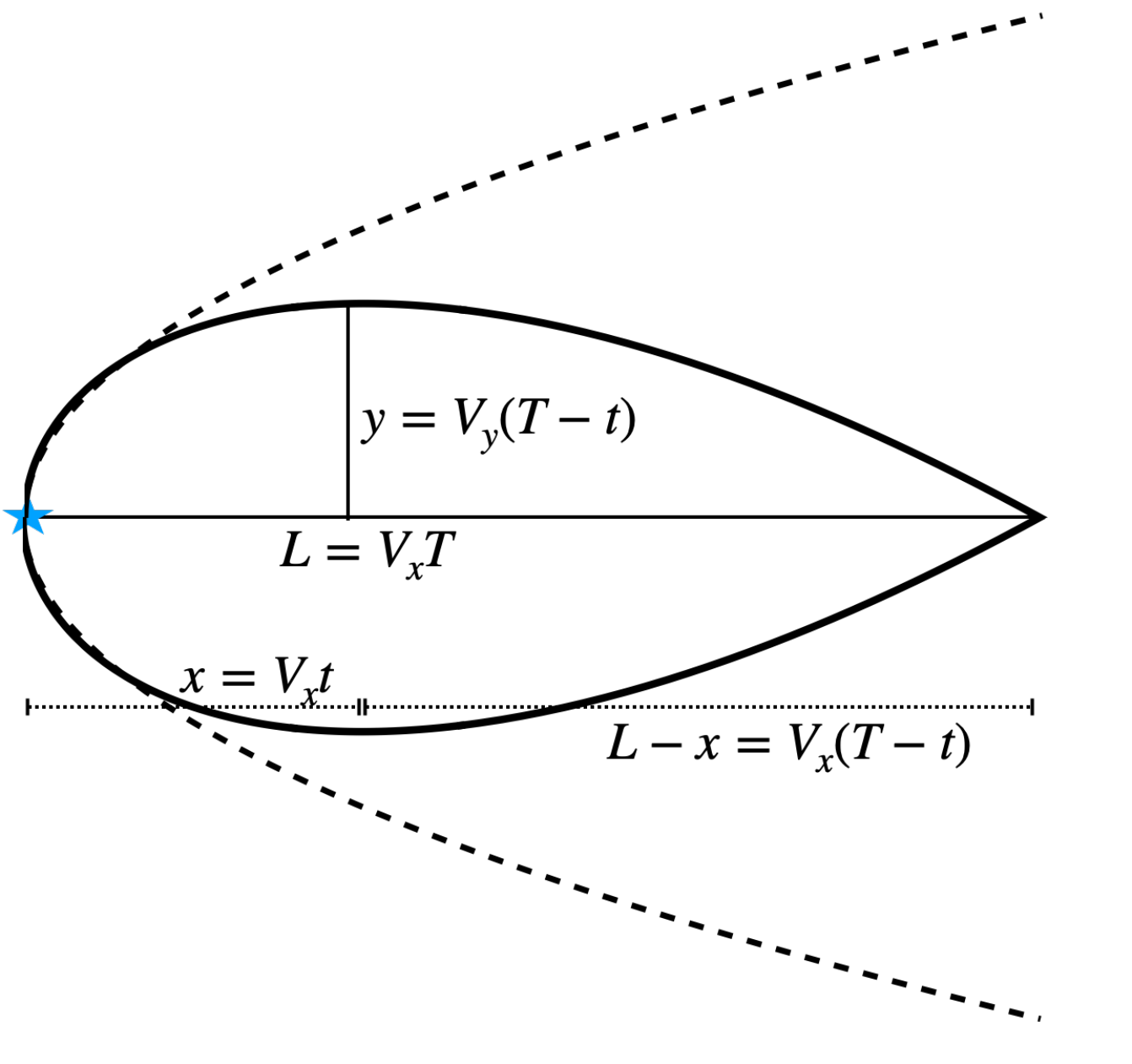}
    \caption{Geometry of the toy model that represents the outflow cavity generated by a collimated high-velocity jet that moves in the $+x$ direction with constant velocity $V_x$ during a time $T$ until it reaches a point $L$. The cavity is expelled axisymetrically in all perpendicular directions from the jet axis. When the jet is in the position $x$ at time $t$ the cavity will travel in the radial direction $+y$ during a time $T-t$ with velocity $V_y$, which is function of the density contrast and therefore, a function of $x$. The locus of the cavity is shown in a thick black line. The dotted line is the expected locus at an infinite time. For a short $x$, the cavity coincide in both cases. }
    \label{fig:OModel}
\end{figure}

To analyze the distribution and kinematics of the \COgas line data, we used an analytic model that can predict the morphology and kinematics of a jet-driven outflow, based on the lateral expansion of the envelope \citep{Shu1991, Raga1993,Rivera19}. We assume a jet with constant velocity $V_j$ moves in the direction $+x$ from the origin. After time $T$ the jet would have reached a length $L=V_j T$. As the jet moves, a part of the envelope will be pushed away due to the thermal pressure produced by the leading shock. 
In this scenario, the jet is traveling in the direction $x$, and there is material ejected 
perpendicularly to the jet direction (i.e.\ in the direction $\pm y$) that creates the outflow cavity.  
We assume that the lateral shock produced at any given point $x<L$ is ballistically ejected with a velocity $V_y(x)$, which is the lateral shock velocity that depends on the density contrast between the ejected material and the envelope. The density is a function of the radial distance from the source $n(x)\propto x^{-\alpha}$, and the density contrast is, in a first approximation a function of $x$,
where $\alpha$ is the density power law index
of the environment.
Thus, a lateral shock ejected at a time $t$ and position $x=V_j t$ would take a time $T-t$ to reach a position $y=V_y(T-t)$. This mechanism is geometrically explained in figure \ref{fig:OModel}.
Additionally, we assume that for any $x$ the material was ejected as a perpendicular wind. This material is ejected at the same velocity, but its expansion velocity $V_y$ is given by the contrast density between the ejected material and the environment. We assume a dependence of $V_y(x)\propto x^{\alpha/2}$, which can be written as $V_y=V_{y0}(x/x_0)^{\alpha/2}$, where $V_{y0}$ is the cavity expansion velocity in a scale $x_0$ close to the source \citep{Shu1991}. Plugging together these assumptions we can obtain the locus for constant velocity jet-driven outflow as
\begin{equation}
    y=\pm \frac{V_{y0}}{V_j}\left( \frac{x}{x_0}\right)^{\alpha/2}(L-x),
    \label{eq:toymodel}
\end{equation}
which is valid for any $ \left| x \right|<L$.

\begin{figure}
    \centering
    \includegraphics[width=\linewidth]{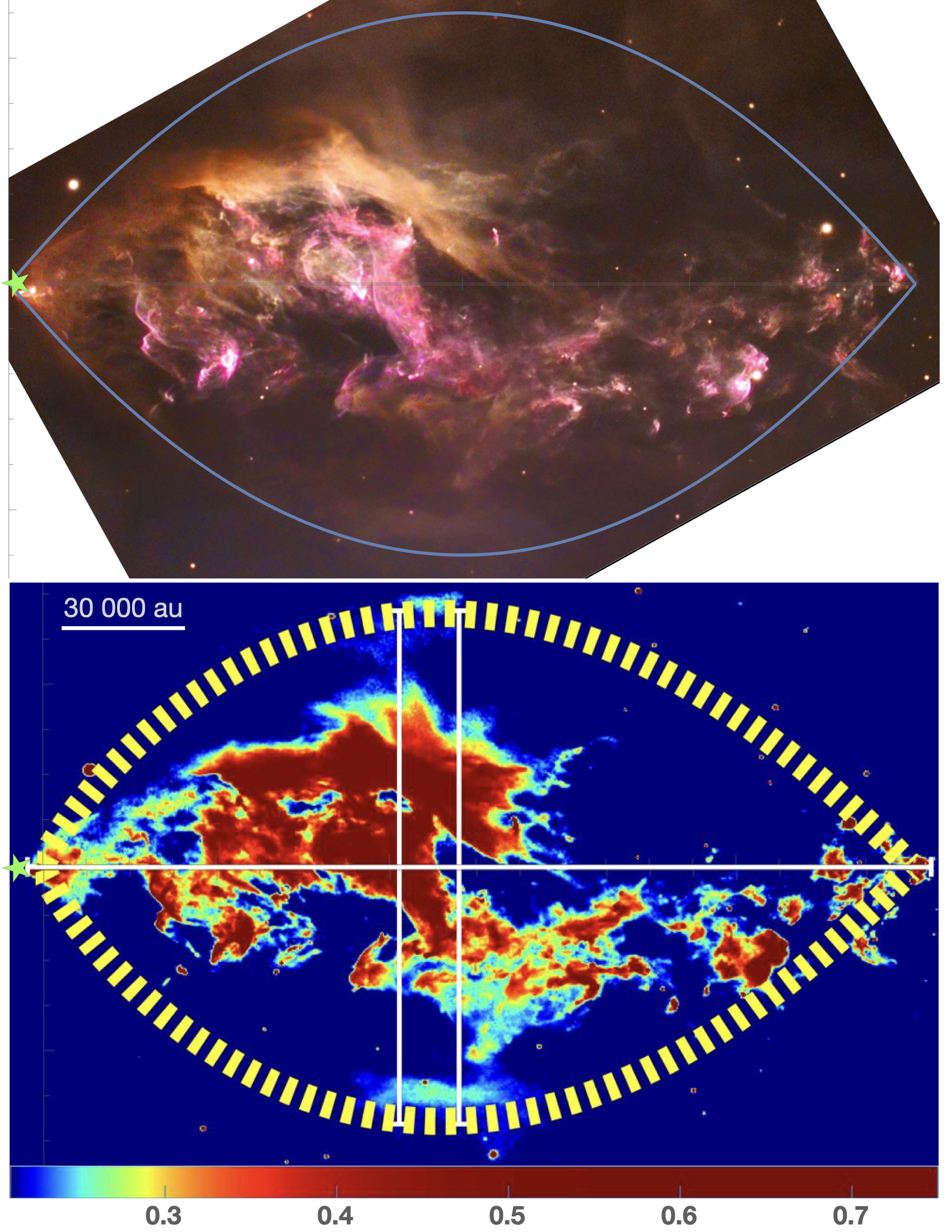}
    \caption{The outflow model (solid line for the upper panel, and dashed line for the lower panel) superimposed to a \Lynds optical image.   Note that the main source is located at the left of the image, represented by the green star. Image captured by Bo Reipurth using the 8m Subaru telescope; color composition by Robert Gendler, using deep H$\alpha$ and [SII] narrow filters.
    The lower panel is a normalized intensity map obtained from the upper panel, using the maximum intensity from the composite image, and all features are above 25\% of the maximum intensity. The white horizontal line is the outflow main axis, and the white vertical lines correspond to the outflow maximum width, which is $y_{max}$, and they cross the main axis at $x_{max}$.}
    \label{fig:physicalModel}
\end{figure}

The curve described by equation \ref{eq:toymodel} is used to create a revolution solid to delimit the outflow surface, the outflow opening angle close to the protostellar source, and its velocity field.  Equation \ref{eq:toymodel} can help us to obtain the maximum outflow width using its derivative, which will be located at 

    \begin{eqnarray}
        {y_{max}}=&{\left(\frac{V_{y0}}{V_{j}}\right)\left(\frac{\alpha L}{x_0(\alpha+2)}\right)^{\alpha/2}\frac{2 L}{\alpha+2}}
        \label{ymax}
        \\
        x_{max}=&\frac{\alpha L}{\alpha+2}.
        \label{eq:xmax}
\end{eqnarray}

Now, using the maximum width of \Lynds in figure \ref{fig:physicalModel}, we find ${y_{max}=(0.34 \pm 0.04)L}$, and this width is enclosed between the projected
$x_{max,p}=(0.45 \pm 0.03)L$,
in terms of the apparent outflow length $L_p$.
From equation \ref{eq:xmax}, we can determine
${\alpha=1.7 \pm 0.2}$,
which is independent of projection effects, since $x_{max,p}/L_p=x_{max}/L$. 
This value determines the density gradient that is associated with the prestellar core and similar to the value considered in models such as the density distribution obtained by \citet{Crimier2010} in the intermediate mass star forming region Cep E region through the dust emission using 1DUSTY, and the model obtained for MCW 349A by \citet{MartinezHenares2023}, who have obtained a value for $\alpha \sim 2$. More specifically, \citet{Shirley2000} has determined a value $\alpha\sim 2.1$ in a survey of 21 low mass cores ranging from prestellar to Class I objects, and \citet{Motte2001} reported $\alpha=2.1\pm0.4$ for \LyndsP, which is consistent with an infalling envelope, that ranges from $\alpha=1.5$ to $2$, depending on the infalling model.  Assuming an internal core size of ${500}$ au, which is about the scale size shown in Figure \ref{fig:PVdisk}, and an outflow length of $L=2\times 10^{5}$ au, the term ${V_j/V_{y0}\sim 100}$. The uncertainty in this case is larger, since this result depends on the angle of the outflow with respect to the plane of the sky because $y_{max}$ does not depend on this angle and $L$ does. More importantly, the assumption for $x_{0}$ would need to be revised. Because of all these assumptions, the value  ${V_j/V_{y0}\sim 100}$ is an estimation of the order of magnitude of this ratio.

However, we can reproduce the morphology and kinematics of the outflow using a simple analytical model. The outflow locus has been used to produce a revolution solid around the jet axis. At the same time, we assume the outflow cavities will expand with a velocity $V_y$ and rotate with a constant circular velocity $V_c$ due to the envelope rotation. Since this is an axisymmetric effect, the outflow morphology is not going to be affected, even when the channel maps could be distorted. This model can be rotated by an angle $\theta$ with respect to the plane perpendicular to the outflow expansion. Even more, the velocity field can be also rotated to obtain its projected component in the direction of an observer, that is, its radial velocity, which is used to produce a synthetic image of the outflow.
We assume that the emission is a function of the gas density. Then the gas density decreases as $r^{-\alpha}$, which means the emission also decreases. We have used $r^{-2}$ to simplify our synthetic model. 
Finally, this analysis can help us to produce a data cube with coordinates position-position-velocity, that can be compared morphologically to the observation, using the position angle of \Lynds to align the synthetic cube to the source observation.

\begin{figure}
    \centering
    \includegraphics[width=\linewidth]{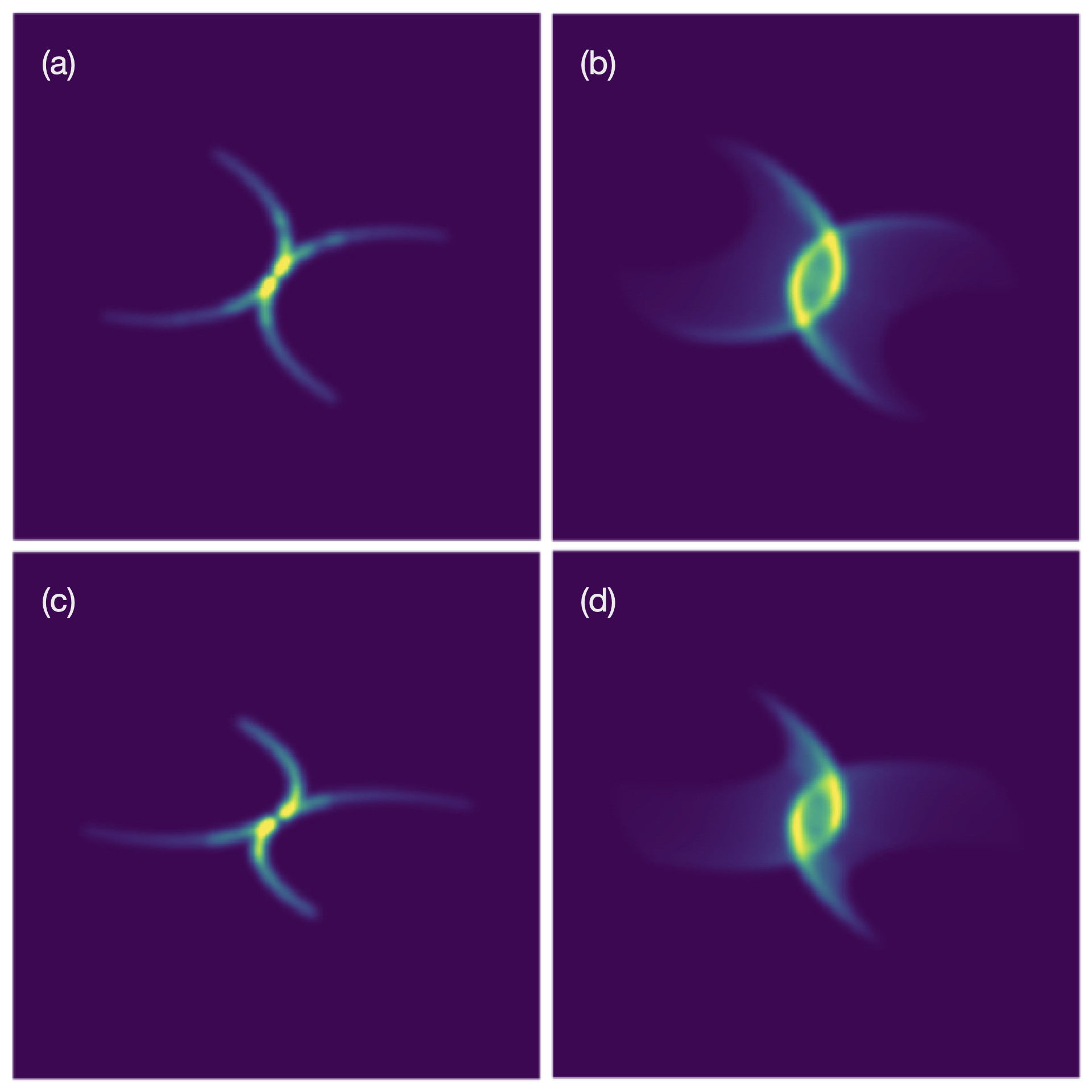}
    \caption{Synthetic cubes with ${V_{y0}=0.1V_{j}}$. Left panels (a and c) represent a single channel map from the analytical model, and right panels (b and d) represent the moment zero from the corresponding model. The upper panels represent a non-rotating model, while the lower panels are models with a $V_c=0.1 V_j$ rotation velocity.    }
    \label{fig:model1}
\end{figure}

\begin{figure}
    \centering
    \includegraphics[width=\linewidth]{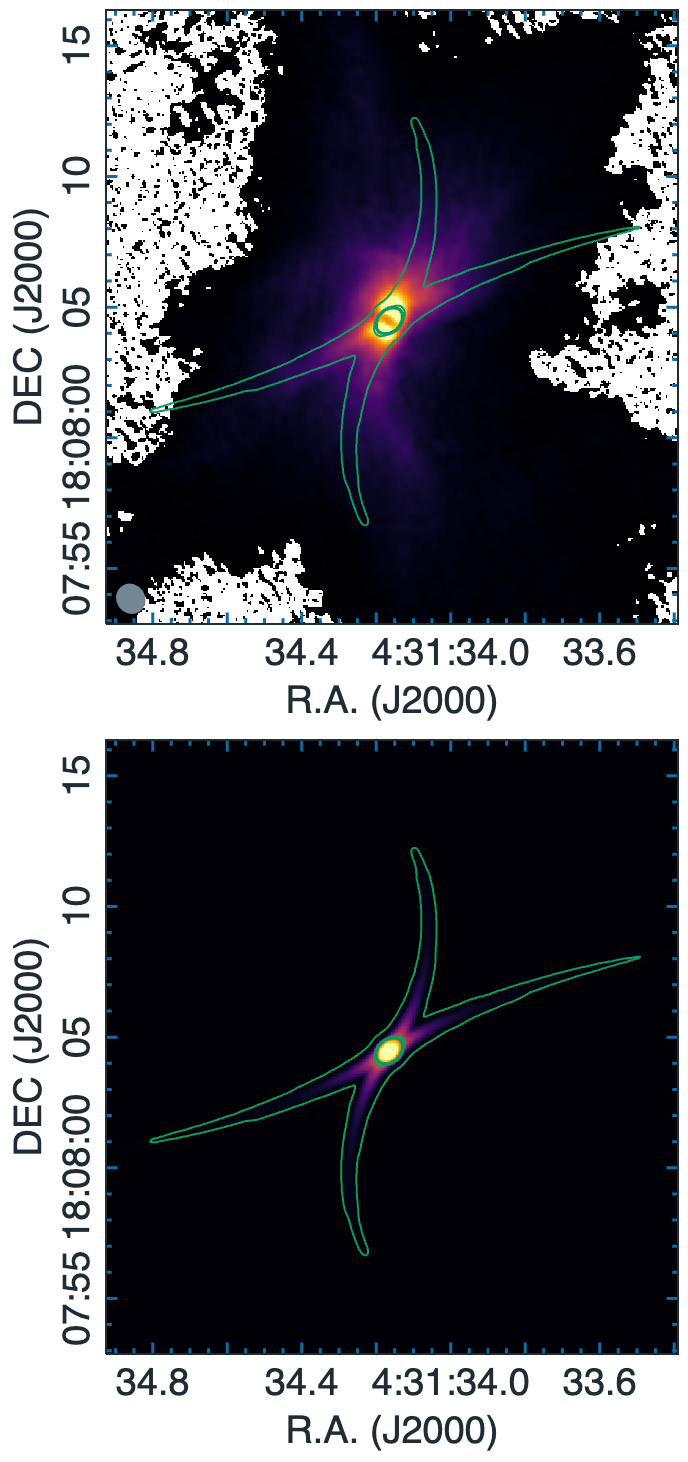}
    \caption{Moment zero contours from our analytic toy model overlaid onto (upper panel): moment zero from the \COgas line observations of \Lynds; (bottom panel): moment zero from the analytical model itself. }
    \label{fig:mom0_obsTmodel}
\end{figure}

To show the distortion, using a set of arbitrary parameters, we produced a set of synthetic cubes, using ${V_{y0}=0.1V_{j}}$. A channel map and a moment zero map from a non-rotating ($V_c=0$) model are presented in Figure \ref{fig:model1} upper panels (a - b), while the lower panels (c - d) show the same set of parameters with an envelope rotation $V_c=0.1V_j$  which exhibits an X-shaped structure that gets an antisymmetry with respect to the ejecting source. 
Therefore, the envelope velocity $V_c$ is a free parameter controlling the level of antisymmetry observed in moment zero maps. Fitting our analytical model to the observation, we have obtained a value of $V_c=2$ km~s$^{-1}$. Including a circular velocity component provides an x-like structure even though we recognize that the rotation velocity model used here is a first approximation since there is expected to be angular momentum in the infalling material which is then entrained with the outflow and pushed outward. 
In the lower panel of  figure \ref{fig:mom0_obsTmodel} 
 we show the moment zero from the model of the outflow for \LyndsP,  and we compare it to the moment zero from the observation with overlapping contours corresponding to the synthetic model's moment zero emission. This comparison shows that our analytic toy model is able to reproduce the main structures and kinematics from this source. We note, however, that reproducing completely the detailed structure that it exhibits will require to incorporate more components in the future.

\section{Conclusions}
\label{sec: Conclusions}
By using Band 6 ALMA observations of the 1.3 mm continuum and \COgas emission, we have analyzed the structure and kinematics of the \Lynds binary system at scales from tens to thousands of au.  We have used different imaging techniques to reveal different structures at disk and envelope scales. We can summarize our results in the following points:

\begin{itemize}
    \item Using the continuum flux we calculated some physical parameters such as the size of the circumbinary disk and the masses for each disk (circumbinary and circumstellar). We show, in particular, that the circumstellar disks have masses of order 1\% of their associated protostar. The circumbinary disk brightness distribution is asymmetric, possibly because of tidal effects caused by the central binary.
        
    \item The \COgas channel maps are dominated by curved filaments mostly associated with the walls of the outflow cavity. The moment zero map reveals the extended structure at envelope scales and further delineates the cavity wall associated with the outflow. The moment one map reveals the rotation of the material. We determined the mass of each component in the system using their \COgas emission.

    \item Using the PV diagrams we updated the specific angular momentum $j$, to  ${270 \pm 30}$ au km s$^{-1}$, and we were able to constrain the central mass of \LyndsP, obtaining a value consistent with previous works that used different methods for mass calculation. 
    The velocity excess in the northern part of the PV diagram may be due to the binary interaction. This could be because the Keplerian motion between the binary components creates a tidal effect in the CBD. This needs to be explored with more numerical models. 
    \item We created a toy model that accounts for the main kinematic components produced by the outflow. It includes the cavity expansion both in the jet direction and the perpendicular direction. Using the outflow morphology, we obtained that the environment is stratified in density with a power law with an index ${\alpha=1.7}$, a value similar to what has previously been found in other protostellar regions.
    \item The synthetic data cube produced with the toy model has been analyzed and compared with the observations, obtaining moment zero and moment one maps, finding that most of the \COgas line is emitted from the cavity walls base, and the rest of the dynamical features could be attributed to infalling material  connecting the envelope and the CBD.
    \item From the asymmetry in the double tunning fork observed in the velocity channels, we obtained the envelope rotation velocity to be $\sim 2$ km s$^{-1}$.
    \item An analytical model is not enough to reproduce the filament-like structures detected in the observations, driving us to use a hydrodynamical model in future papers to prove the extent of the binary influence in this architecture.

\end{itemize}

\section*{Acknowledgements}
We thank the anonymous referee for their comments to improve the clarity of this paper. 
L.L. acknowledges the support of UNAM-DGAPA PAPIIT grants IN112820 and IN108324, and CONACYT-CF grant 263356. P.R-O. acknowledges UNAM-PAPIIT grants IN110722, IN103921, IN113119, IG100422, CONACYT grant 280775.
ClCo, LP and GS acknowledge the PRIN-MUR 2020  BEYOND-2p (Astrochemistry beyond the second period elements, Prot. 2020AFB3FX), the project ASI-Astrobiologia 2023 MIGLIORA (Modeling Chemical Complexity, F83C23000800005), the INAF-GO 2023 fundings PROTO-SKA (Exploiting ALMA data to study planet forming disks: preparing the advent of SKA, C13C23000770005), the INAF Mini-Grant 2023 TRIESTE (``TRacing the chemIcal hEritage of our originS: from proTostars to planEts''; PI: G. Sabatini),
and the National Recovery and Resilience Plan (NRRP), Mission 4, Component 2, Investment 1.1, Call for tender No. 104 published on 2.2.2022 by the Italian Ministry of University and Research (MUR), funded by the European Union – NextGenerationEU– Project Title 2022JC2Y93 Chemical Origins: linking the fossil composition of the Solar System with the chemistry of protoplanetary disks – CUP J53D23001600006 - Grant Assignment Decree No. 962 adopted on 30.06.2023 by the Italian Ministry of Ministry of University and Research (MUR). This project has received funding from the European Research Council (ERC) under the European Union Horizon Europe programme (grant agreement No. 101042275, project Stellar-MADE). This paper makes use of the following ALMA data: ADS/JAO.2018.1.01205.L. ALMA is a partnership of ESO (representing its member states), NSF (USA) and NINS (Japan), together with NRC
(Canada), NSC and ASIAA (Taiwan), and KASI (Republic of Korea), in cooperation with the Republic of Chile. The Joint ALMA Observatory is operated by ESO, AUI/NRAO and NAOJ. E.B. acknowledges the contribution of the Next Generation EU funds within the National Recovery and Resilience Plan (PNRR), Mission 4 - Education and Research, Component 2 - From Research to Business (M4C2), Investment Line 3.1 - Strengthening and creation of Research Infrastructures, Project IR0000034 – STILES - Strengthening the Italian Leadership in ELT and SKA.

\section*{Data Availability}
The data underlying this article will be shared on reasonable request to the corresponding author.

\bibliographystyle{mnras}
\bibliography{L1551_article} 

\begin{thebibliography}{}
\makeatletter
\relax
\def\mn@urlcharsother{\let\do\@makeother \do\$\do\&\do\#\do\^\do\_\do\%\do\~}
\def\mn@doi{\begingroup\mn@urlcharsother \@ifnextchar [ {\mn@doi@} {\mn@doi@[]}}
\def\mn@doi@[#1]#2{\def\@tempa{#1}\ifx\@tempa\@empty \href {http://dx.doi.org/#2} {doi:#2}\else \href {http://dx.doi.org/#2} {#1}\fi \endgroup}
\def\mn@eprint#1#2{\mn@eprint@#1:#2::\@nil}
\def\mn@eprint@arXiv#1{\href {http://arxiv.org/abs/#1} {{\tt arXiv:#1}}}
\def\mn@eprint@dblp#1{\href {http://dblp.uni-trier.de/rec/bibtex/#1.xml} {dblp:#1}}
\def\mn@eprint@#1:#2:#3:#4\@nil{\def\@tempa {#1}\def\@tempb {#2}\def\@tempc {#3}\ifx \@tempc \@empty \let \@tempc \@tempb \let \@tempb \@tempa \fi \ifx \@tempb \@empty \def\@tempb {arXiv}\fi \@ifundefined {mn@eprint@\@tempb}{\@tempb:\@tempc}{\expandafter \expandafter \csname mn@eprint@\@tempb\endcsname \expandafter{\@tempc}}}

\bibitem[\protect\citeauthoryear{{Adams}, {Lada}  \& {Shu}}{{Adams} et~al.}{1987}]{Adams1987}
{Adams} F.~C.,  {Lada} C.~J.,   {Shu} F.~H.,  1987, \mn@doi [\apj] {10.1086/164924}, \href {https://ui.adsabs.harvard.edu/abs/1987ApJ...312..788A} {312, 788}

\bibitem[\protect\citeauthoryear{{Adams}, {Ruden}  \& {Shu}}{{Adams} et~al.}{1989}]{Adams1989}
{Adams} F.~C.,  {Ruden} S.~P.,   {Shu} F.~H.,  1989, \mn@doi [\apj] {10.1086/168187}, \href {https://ui.adsabs.harvard.edu/abs/1989ApJ...347..959A} {347, 959}

\bibitem[\protect\citeauthoryear{{Andr{\'e}}, {Motte}, {Bacmann}  \& {Belloche}}{{Andr{\'e}} et~al.}{1999}]{Andre1999}
{Andr{\'e}} P.,  {Motte} F.,  {Bacmann} A.,   {Belloche} A.,  1999, in {Nakamoto} T.,  ed., Star Formation 1999. pp 145--152

\bibitem[\protect\citeauthoryear{{Andre}, {Ward-Thompson}  \& {Barsony}}{{Andre} et~al.}{2000}]{Andre2000}
{Andre} P.,  {Ward-Thompson} D.,   {Barsony} M.,  2000, in {Mannings} V.,  {Boss} A.~P.,   {Russell} S.~S.,  eds, Protostars and Planets IV. p.~59 (\mn@eprint {arXiv} {astro-ph/9903284}), \mn@doi{10.48550/arXiv.astro-ph/9903284}

\bibitem[\protect\citeauthoryear{{Ansdell}, {Williams}, {Manara}, {Miotello}, {Facchini}, {van der Marel}, {Testi}  \& {van Dishoeck}}{{Ansdell} et~al.}{2017}]{Ansdell2017}
{Ansdell} M.,  {Williams} J.~P.,  {Manara} C.~F.,  {Miotello} A.,  {Facchini} S.,  {van der Marel} N.,  {Testi} L.,   {van Dishoeck} E.~F.,  2017, \mn@doi [\aj] {10.3847/1538-3881/aa69c0}, \href {https://ui.adsabs.harvard.edu/abs/2017AJ....153..240A} {153, 240}

\bibitem[\protect\citeauthoryear{{Bergin} \& {Williams}}{{Bergin} \& {Williams}}{2017}]{Bergin2017}
{Bergin} E.~A.,  {Williams} J.~P.,  2017, in {Pessah} M.,  {Gressel} O.,  eds,  Astrophysics and Space Science Library Vol. 445, Formation, Evolution, and Dynamics of Young Solar Systems. p.~1, \mn@doi{10.1007/978-3-319-60609-5_1}

\bibitem[\protect\citeauthoryear{{Bianchi} et~al.,}{{Bianchi} et~al.}{2020}]{Bianchi2020}
{Bianchi} E.,  et~al., 2020, \mn@doi [MNRAS] {10.1093/mnrasl/slaa130}, \href {https://ui.adsabs.harvard.edu/abs/2020MNRAS.tmpL.141B} {}

\bibitem[\protect\citeauthoryear{{Bohlin}, {Savage}  \& {Drake}}{{Bohlin} et~al.}{1978}]{Bohlin1978}
{Bohlin} R.~C.,  {Savage} B.~D.,   {Drake} J.~F.,  1978, \mn@doi [\apj] {10.1086/156357}, \href {https://ui.adsabs.harvard.edu/abs/1978ApJ...224..132B} {224, 132}

\bibitem[\protect\citeauthoryear{{Chou}, {Takakuwa}, {Yen}, {Ohashi}  \& {Ho}}{{Chou} et~al.}{2014}]{Chou2014}
{Chou} T.-L.,  {Takakuwa} S.,  {Yen} H.-W.,  {Ohashi} N.,   {Ho} P. T.~P.,  2014, \mn@doi [\apj] {10.1088/0004-637X/796/1/70}, \href {https://ui.adsabs.harvard.edu/abs/2014ApJ...796...70C} {796, 70}

\bibitem[\protect\citeauthoryear{{Codella}, {Ceccarelli}, {Chandler}, {Sakai}, {Yamamoto}  \& {FAUST Team}}{{Codella} et~al.}{2021}]{Codella2021}
{Codella} C.,  {Ceccarelli} C.,  {Chandler} C.,  {Sakai} N.,  {Yamamoto} S.,   {FAUST Team} 2021, \mn@doi [Frontiers in Astronomy and Space Sciences] {10.3389/fspas.2021.782006}, \href {https://ui.adsabs.harvard.edu/abs/2021FrASS...8..227C} {8, 227}

\bibitem[\protect\citeauthoryear{{Codella} et~al.,}{{Codella} et~al.}{2024}]{Codella2024}
{Codella} C.,  et~al., 2024, \mn@doi [\mnras] {10.1093/mnras/stae472}, \href {https://ui.adsabs.harvard.edu/abs/2024MNRAS.528.7383C} {528, 7383}

\bibitem[\protect\citeauthoryear{{Connelley} \& {Reipurth}}{{Connelley} \& {Reipurth}}{2018}]{Connelley2018}
{Connelley} M.~S.,  {Reipurth} B.,  2018, \mn@doi [\apj] {10.3847/1538-4357/aaba7b}, \href {https://ui.adsabs.harvard.edu/abs/2018ApJ...861..145C} {861, 145}

\bibitem[\protect\citeauthoryear{{Cortes-Rangel}, {Zapata}, {Toal{\'a}}, {Ho}, {Takahashi}, {Mesa-Delgado}  \& {Masqu{\'e}}}{{Cortes-Rangel} et~al.}{2020}]{Cortes-Rangel2020}
{Cortes-Rangel} G.,  {Zapata} L.~A.,  {Toal{\'a}} J.~A.,  {Ho} P. T.~P.,  {Takahashi} S.,  {Mesa-Delgado} A.,   {Masqu{\'e}} J.~M.,  2020, \mn@doi [\aj] {10.3847/1538-3881/ab6295}, \href {https://ui.adsabs.harvard.edu/abs/2020AJ....159...62C} {159, 62}

\bibitem[\protect\citeauthoryear{{Crimier} et~al.,}{{Crimier} et~al.}{2010}]{Crimier2010}
{Crimier} N.,  et~al., 2010, \mn@doi [\aap] {10.1051/0004-6361/200913499}, \href {https://ui.adsabs.harvard.edu/abs/2010A&A...516A.102C} {516, A102}

\bibitem[\protect\citeauthoryear{{Cruz-S{\'a}enz de Miera}, {K{\'o}sp{\'a}l}, {{\'A}brah{\'a}m}, {Liu}  \& {Takami}}{{Cruz-S{\'a}enz de Miera} et~al.}{2019}]{Cruz2019}
{Cruz-S{\'a}enz de Miera} F.,  {K{\'o}sp{\'a}l} {\'A}.,  {{\'A}brah{\'a}m} P.,  {Liu} H.~B.,   {Takami} M.,  2019, \mn@doi [ApJL] {10.3847/2041-8213/ab39ea}, \href {https://ui.adsabs.harvard.edu/abs/2019ApJ...882L...4C} {882, L4}

\bibitem[\protect\citeauthoryear{{Duch{\^e}ne}, {Bontemps}, {Bouvier}, {Andr{\'e}}, {Djupvik}  \& {Ghez}}{{Duch{\^e}ne} et~al.}{2007}]{Duchene2007}
{Duch{\^e}ne} G.,  {Bontemps} S.,  {Bouvier} J.,  {Andr{\'e}} P.,  {Djupvik} A.~A.,   {Ghez} A.~M.,  2007, \mn@doi [\aap] {10.1051/0004-6361:20077270}, \href {https://ui.adsabs.harvard.edu/abs/2007A&A...476..229D} {476, 229}

\bibitem[\protect\citeauthoryear{{Feeney-Johansson} et~al.,}{{Feeney-Johansson} et~al.}{2023}]{Feeney-Johansson2023}
{Feeney-Johansson} A.,  et~al., 2023, \mn@doi [\aap] {10.1051/0004-6361/202346737}, \href {https://ui.adsabs.harvard.edu/abs/2023A&A...677A..97F} {677, A97}

\bibitem[\protect\citeauthoryear{{Frank} et~al.,}{{Frank} et~al.}{2014}]{Frank2014}
{Frank} A.,  et~al., 2014, in {Beuther} H.,  {Klessen} R.~S.,  {Dullemond} C.~P.,   {Henning} T.,  eds, Protostars and Planets VI. pp 451--474 (\mn@eprint {arXiv} {1402.3553}), \mn@doi{10.2458/azu_uapress_9780816531240-ch020}

\bibitem[\protect\citeauthoryear{{Galli} et~al.,}{{Galli} et~al.}{2019}]{Galli2019}
{Galli} P.~A.~B.,  et~al., 2019, \mn@doi [\aap] {10.1051/0004-6361/201935928}, \href {https://ui.adsabs.harvard.edu/abs/2019A&A...630A.137G} {630, A137}

\bibitem[\protect\citeauthoryear{{Goodwin}, {Whitworth}  \& {Ward-Thompson}}{{Goodwin} et~al.}{2004}]{Goodwin2004}
{Goodwin} S.~P.,  {Whitworth} A.~P.,   {Ward-Thompson} D.,  2004, \mn@doi [\aap] {10.1051/0004-6361:20031594}, \href {https://ui.adsabs.harvard.edu/abs/2004A&A...414..633G} {414, 633}

\bibitem[\protect\citeauthoryear{{Hern{\'a}ndez Garnica} et~al.,}{{Hern{\'a}ndez Garnica} et~al.}{2024}]{HernandezGarnica2024}
{Hern{\'a}ndez Garnica} R.,  et~al., 2024, \mn@doi [\mnras] {10.1093/mnras/stae2482}, \href {https://ui.adsabs.harvard.edu/abs/2024MNRAS.535.2948H} {535, 2948}

\bibitem[\protect\citeauthoryear{{Lim}, {Yeung}, {Hanawa}, {Takakuwa}, {Matsumoto}  \& {Saigo}}{{Lim} et~al.}{2016}]{Lim2016}
{Lim} J.,  {Yeung} P. K.~H.,  {Hanawa} T.,  {Takakuwa} S.,  {Matsumoto} T.,   {Saigo} K.,  2016, \mn@doi [ApJ] {10.3847/0004-637X/826/2/153}, \href {https://ui.adsabs.harvard.edu/abs/2016ApJ...826..153L} {826, 153}

\bibitem[\protect\citeauthoryear{{Liseau}, {Fridlund}  \& {Larsson}}{{Liseau} et~al.}{2005}]{Liseau2005}
{Liseau} R.,  {Fridlund} C.~V.~M.,   {Larsson} B.,  2005, \mn@doi [\apj] {10.1086/426783}, \href {https://ui.adsabs.harvard.edu/abs/2005ApJ...619..959L} {619, 959}

\bibitem[\protect\citeauthoryear{{Looney}, {Mundy}  \& {Welch}}{{Looney} et~al.}{1997}]{Looney1997}
{Looney} L.~W.,  {Mundy} L.~G.,   {Welch} W.~J.,  1997, \mn@doi [\apjl] {10.1086/310795}, \href {https://ui.adsabs.harvard.edu/abs/1997ApJ...484L.157L} {484, L157}

\bibitem[\protect\citeauthoryear{{Mart{\'\i}nez-Henares} et~al.,}{{Mart{\'\i}nez-Henares} et~al.}{2023}]{MartinezHenares2023}
{Mart{\'\i}nez-Henares} A.,  et~al., 2023, \mn@doi [\apj] {10.3847/1538-4357/acebcd}, \href {https://ui.adsabs.harvard.edu/abs/2023ApJ...955..119M} {955, 119}

\bibitem[\protect\citeauthoryear{{Maureira}, {Pineda}, {Segura-Cox}, {Caselli}, {Testi}, {Lodato}, {Loinard}  \& {Hern{\'a}ndez-G{\'o}mez}}{{Maureira} et~al.}{2020}]{Maureira2020}
{Maureira} M.~J.,  {Pineda} J.~E.,  {Segura-Cox} D.~M.,  {Caselli} P.,  {Testi} L.,  {Lodato} G.,  {Loinard} L.,   {Hern{\'a}ndez-G{\'o}mez} A.,  2020, \mn@doi [\apj] {10.3847/1538-4357/ab960b}, \href {https://ui.adsabs.harvard.edu/abs/2020ApJ...897...59M} {897, 59}

\bibitem[\protect\citeauthoryear{{McKee} \& {Ostriker}}{{McKee} \& {Ostriker}}{2007}]{McKee2007}
{McKee} C.~F.,  {Ostriker} E.~C.,  2007, \mn@doi [\araa] {10.1146/annurev.astro.45.051806.110602}, \href {https://ui.adsabs.harvard.edu/abs/2007ARA&A..45..565M} {45, 565}

\bibitem[\protect\citeauthoryear{{Momose}, {Ohashi}, {Kawabe}, {Nakano}  \& {Hayashi}}{{Momose} et~al.}{1998}]{Momose1998}
{Momose} M.,  {Ohashi} N.,  {Kawabe} R.,  {Nakano} T.,   {Hayashi} M.,  1998, \mn@doi [\apj] {10.1086/306061}, \href {https://ui.adsabs.harvard.edu/abs/1998ApJ...504..314M} {504, 314}

\bibitem[\protect\citeauthoryear{{Motte} \& {Andr{\'e}}}{{Motte} \& {Andr{\'e}}}{2001}]{Motte2001}
{Motte} F.,  {Andr{\'e}} P.,  2001, \mn@doi [\aap] {10.1051/0004-6361:20000072}, \href {https://ui.adsabs.harvard.edu/abs/2001A&A...365..440M} {365, 440}

\bibitem[\protect\citeauthoryear{{Mundt}, {Stocke}, {Strom}, {Strom}  \& {Anderson}}{{Mundt} et~al.}{1985}]{Mundt1985}
{Mundt} R.,  {Stocke} J.,  {Strom} S.~E.,  {Strom} K.~M.,   {Anderson} E.~R.,  1985, \mn@doi [\apjl] {10.1086/184554}, \href {https://ui.adsabs.harvard.edu/abs/1985ApJ...297L..41M} {297, L41}

\bibitem[\protect\citeauthoryear{{Okuzumi}}{{Okuzumi}}{2025}]{Okuzumi2025}
{Okuzumi} S.,  2025, \mn@doi [\pasj] {10.1093/pasj/psae107}, \href {https://ui.adsabs.harvard.edu/abs/2025PASJ...77..162O} {77, 162}

\bibitem[\protect\citeauthoryear{Oya}{Oya}{2022}]{Oya2022}
Oya Y.,  2022, Model Calculation.
Springer Nature Singapore, Singapore, pp 31--50, \mn@doi{10.1007/978-981-19-1708-0_3}, \url {https://doi.org/10.1007/978-981-19-1708-0_3}

\bibitem[\protect\citeauthoryear{{Oya}, {Sakai}, {L{\'o}pez-Sepulcre}, {Watanabe}, {Ceccarelli}, {Lefloch}, {Favre}  \& {Yamamoto}}{{Oya} et~al.}{2016}]{Oya2016}
{Oya} Y.,  {Sakai} N.,  {L{\'o}pez-Sepulcre} A.,  {Watanabe} Y.,  {Ceccarelli} C.,  {Lefloch} B.,  {Favre} C.,   {Yamamoto} S.,  2016, \mn@doi [\apj] {10.3847/0004-637X/824/2/88}, \href {https://ui.adsabs.harvard.edu/abs/2016ApJ...824...88O} {824, 88}

\bibitem[\protect\citeauthoryear{{Padoan}, {Nordlund}, {Kritsuk}, {Norman}  \& {Li}}{{Padoan} et~al.}{2007}]{Padoan2007}
{Padoan} P.,  {Nordlund} {\r{A}}.,  {Kritsuk} A.~G.,  {Norman} M.~L.,   {Li} P.~S.,  2007, \mn@doi [\apj] {10.1086/516623}, \href {https://ui.adsabs.harvard.edu/abs/2007ApJ...661..972P} {661, 972}

\bibitem[\protect\citeauthoryear{{Park} et~al.,}{{Park} et~al.}{2021}]{Park2021}
{Park} W.,  et~al., 2021, \mn@doi [\apj] {10.3847/1538-4357/ac1745}, \href {https://ui.adsabs.harvard.edu/abs/2021ApJ...920..132P} {920, 132}

\bibitem[\protect\citeauthoryear{{Pineda} et~al.,}{{Pineda} et~al.}{2023}]{Pineda2023}
{Pineda} J.~E.,  et~al., 2023, in {Inutsuka} S.,  {Aikawa} Y.,  {Muto} T.,  {Tomida} K.,   {Tamura} M.,  eds,  Astronomical Society of the Pacific Conference Series Vol. 534, Protostars and Planets VII. p.~233 (\mn@eprint {arXiv} {2205.03935}), \mn@doi{10.48550/arXiv.2205.03935}

\bibitem[\protect\citeauthoryear{{Raga} \& {Cabrit}}{{Raga} \& {Cabrit}}{1993}]{Raga1993}
{Raga} A.,  {Cabrit} S.,  1993, \aap, \href {https://ui.adsabs.harvard.edu/abs/1993A&A...278..267R} {278, 267}

\bibitem[\protect\citeauthoryear{{Reipurth}, {Clarke}, {Boss}, {Goodwin}, {Rodr{\'\i}guez}, {Stassun}, {Tokovinin}  \& {Zinnecker}}{{Reipurth} et~al.}{2014}]{Reipurth2014}
{Reipurth} B.,  {Clarke} C.~J.,  {Boss} A.~P.,  {Goodwin} S.~P.,  {Rodr{\'\i}guez} L.~F.,  {Stassun} K.~G.,  {Tokovinin} A.,   {Zinnecker} H.,  2014, in {Beuther} H.,  {Klessen} R.~S.,  {Dullemond} C.~P.,   {Henning} T.,  eds, Protostars and Planets VI. p.~267 (\mn@eprint {arXiv} {1403.1907}), \mn@doi{10.2458/azu\_uapress\_9780816531240-ch012}

\bibitem[\protect\citeauthoryear{{Rivera-Ortiz}, {Rodr{\'\i}guez-Gonz{\'a}lez}, {Hern{\'a}ndez-Mart{\'\i}nez}  \& {Cant{\'o}}}{{Rivera-Ortiz} et~al.}{2019}]{Rivera19}
{Rivera-Ortiz} P.~R.,  {Rodr{\'\i}guez-Gonz{\'a}lez} A.,  {Hern{\'a}ndez-Mart{\'\i}nez} L.,   {Cant{\'o}} J.,  2019, \mn@doi [\apj] {10.3847/1538-4357/ab05ca}, \href {https://ui.adsabs.harvard.edu/abs/2019ApJ...874...38R} {874, 38}

\bibitem[\protect\citeauthoryear{{Rodriguez}, {Canto}, {Torrelles}  \& {Ho}}{{Rodriguez} et~al.}{1986}]{Rodriguez1986}
{Rodriguez} L.~F.,  {Canto} J.,  {Torrelles} J.~M.,   {Ho} P.~T.~P.,  1986, \mn@doi [\apjl] {10.1086/184616}, \href {https://ui.adsabs.harvard.edu/abs/1986ApJ...301L..25R} {301, L25}

\bibitem[\protect\citeauthoryear{{Rodr{\'\i}guez}, {Curiel}, {Cant{\'o}}, {Loinard}, {Raga}  \& {Torrelles}}{{Rodr{\'\i}guez} et~al.}{2003a}]{Rodriguez2003a}
{Rodr{\'\i}guez} L.~F.,  {Curiel} S.,  {Cant{\'o}} J.,  {Loinard} L.,  {Raga} A.~C.,   {Torrelles} J.~M.,  2003a, \mn@doi [ApJ] {10.1086/344833}, \href {https://ui.adsabs.harvard.edu/abs/2003ApJ...583..330R} {583, 330}

\bibitem[\protect\citeauthoryear{{Rodr{\'\i}guez}, {Porras}, {Claussen}, {Curiel}, {Wilner}  \& {Ho}}{{Rodr{\'\i}guez} et~al.}{2003b}]{Rodriguez2003b}
{Rodr{\'\i}guez} L.~F.,  {Porras} A.,  {Claussen} M.~J.,  {Curiel} S.,  {Wilner} D.~J.,   {Ho} P. T.~P.,  2003b, \mn@doi [\apjl] {10.1086/374882}, \href {https://ui.adsabs.harvard.edu/abs/2003ApJ...586L.137R} {586, L137}

\bibitem[\protect\citeauthoryear{{Roueff} et~al.,}{{Roueff} et~al.}{2021}]{AntoineRoueff2021}
{Roueff} A.,  et~al., 2021, \mn@doi [\aap] {10.1051/0004-6361/202037776}, \href {https://ui.adsabs.harvard.edu/abs/2021A&A...645A..26R} {645, A26}

\bibitem[\protect\citeauthoryear{{Sakai} et~al.,}{{Sakai} et~al.}{2014}]{Sakai2014}
{Sakai} N.,  et~al., 2014, \mn@doi [\nat] {10.1038/nature13000}, \href {https://ui.adsabs.harvard.edu/abs/2014Natur.507...78S} {507, 78}

\bibitem[\protect\citeauthoryear{{Shirley}, {Evans}, {Rawlings}  \& {Gregersen}}{{Shirley} et~al.}{2000}]{Shirley2000}
{Shirley} Y.~L.,  {Evans} II N.~J.,  {Rawlings} J. M.~C.,   {Gregersen} E.~M.,  2000, \mn@doi [\apjs] {10.1086/317358}, \href {https://ui.adsabs.harvard.edu/abs/2000ApJS..131..249S} {131, 249}

\bibitem[\protect\citeauthoryear{{Shu}, {Adams}  \& {Lizano}}{{Shu} et~al.}{1987}]{Shu1987}
{Shu} F.~H.,  {Adams} F.~C.,   {Lizano} S.,  1987, \mn@doi [\araa] {10.1146/annurev.aa.25.090187.000323}, \href {https://ui.adsabs.harvard.edu/abs/1987ARA&A..25...23S} {25, 23}

\bibitem[\protect\citeauthoryear{{Shu}, {Ruden}, {Lada}  \& {Lizano}}{{Shu} et~al.}{1991}]{Shu1991}
{Shu} F.~H.,  {Ruden} S.~P.,  {Lada} C.~J.,   {Lizano} S.,  1991, \mn@doi [\apjl] {10.1086/185970}, \href {https://ui.adsabs.harvard.edu/abs/1991ApJ...370L..31S} {370, L31}

\bibitem[\protect\citeauthoryear{{Snell}, {Loren}  \& {Plambeck}}{{Snell} et~al.}{1980}]{Snell1980}
{Snell} R.~L.,  {Loren} R.~B.,   {Plambeck} R.~L.,  1980, \mn@doi [\apjl] {10.1086/183283}, \href {https://ui.adsabs.harvard.edu/abs/1980ApJ...239L..17S} {239, L17}

\bibitem[\protect\citeauthoryear{{Stahler}, {Korycansky}, {Brothers}  \& {Touma}}{{Stahler} et~al.}{1994}]{Stahler1994}
{Stahler} S.~W.,  {Korycansky} D.~G.,  {Brothers} M.~J.,   {Touma} J.,  1994, \mn@doi [\apj] {10.1086/174489}, \href {https://ui.adsabs.harvard.edu/abs/1994ApJ...431..341S} {431, 341}

\bibitem[\protect\citeauthoryear{Team et~al.,}{Team et~al.}{2022}]{The_CASA_Team_2022}
Team T.~C.,  et~al., 2022, \mn@doi [Publications of the Astronomical Society of the Pacific] {10.1088/1538-3873/ac9642}, 134, 114501

\bibitem[\protect\citeauthoryear{{Tobin} et~al.,}{{Tobin} et~al.}{2016}]{Tobin2016}
{Tobin} J.~J.,  et~al., 2016, \mn@doi [\apj] {10.3847/0004-637X/818/1/73}, \href {https://ui.adsabs.harvard.edu/abs/2016ApJ...818...73T} {818, 73}

\bibitem[\protect\citeauthoryear{{Vastel} et~al.,}{{Vastel} et~al.}{2022}]{Vastel2022}
{Vastel} C.,  et~al., 2022, \mn@doi [\aap] {10.1051/0004-6361/202243414}, \href {https://ui.adsabs.harvard.edu/abs/2022A&A...664A.171V} {664, A171}

\makeatother
\end{thebibliography}








\bsp	
\label{lastpage}
\end{document}